\documentclass[11pt, a4paper]{article}

\usepackage{graphicx}  
\usepackage{booktabs}  
\usepackage{adjustbox} 
\usepackage{multirow}
\usepackage{rotating}
\usepackage{setspace} 
\usepackage[margin=1in]{geometry} 
\usepackage{pgfplots}
\usepackage{pbox} 
\usepackage{pgfplotstable}
\usepackage{enumitem}   
\usepackage{array}  
\pgfplotsset{compat=1.17}

\usepackage{comment}
\usepackage{tikz}
\usetikzlibrary{shadows}
\usetikzlibrary{shapes.geometric, positioning, arrows.meta, backgrounds, calc, fit}
\usepackage{subcaption}
\usepackage{amsmath, yhmath} 
\usepackage{amssymb}  
\usepackage{tabularx} 
\usepackage{makecell} 
\usepackage[round,authoryear]{natbib}
\bibpunct{(}{)}{;}{a}{,}{,}
\setcitestyle{etal}

\usepackage{hyperref}
\usepackage{threeparttable}
\hypersetup{
  colorlinks   = true,
  urlcolor     = blue,
  linkcolor    = blue,
  citecolor    = blue
}
\usepackage{float}

\makeatletter

\newcommand{\decisionheading}[2]{%
  \section*{Decision #1: #2}%
  \label{decision:#1}%
  \def\@currentlabelname{Decision #1}%
}
\makeatother

\usepackage{authblk}
\providecommand{\keywords}[1]{\textbf{\textit{Keywords:}} #1}

\title{Responsible Innovation: A Strategic Framework for Financial LLM Integration}
\author{
  Ahmadreza Tavasoli, Maedeh Sharbaf, Seyed Mohamad Madani%
  \thanks{Ahmadreza Tavasoli: \texttt{ahmadreza.tavasoli@hec.ca}; 
          Maedeh Sharbaf: \texttt{maedeh90@mit.edu}; 
          Seyed Mohamad Madani: \texttt{m.madani@math.iut.ac.ir}}
}

\date{}

\begin{document}
\maketitle

\begin{abstract}

Financial institutions of all sizes are increasingly adopting Large Language Models (LLMs) to enhance credit assessments, deliver personalized client advisory services, and automate various language-intensive processes. However, effectively deploying LLMs requires careful management of stringent data governance requirements, heightened demands for interpretability, ethical responsibilities, and rapidly evolving regulatory landscapes. To address these challenges, we introduce a structured six-decision framework specifically designed for the financial sector, guiding organizations systematically from initial feasibility assessments to final deployment strategies.

The framework encourages institutions to: (1) evaluate whether an advanced LLM is necessary at all, (2) formalize robust data governance and privacy safeguards, (3) establish targeted risk management mechanisms, (4) integrate ethical considerations early in the development process, (5) justify the initiative's return on investment (ROI) and strategic value, and only then (6) choose the optimal implementation pathway—open-source versus proprietary, or in-house versus vendor-supported—aligned with regulatory requirements and operational realities. By linking strategic considerations with practical steps such as pilot testing, maintaining comprehensive audit trails, and conducting ongoing compliance evaluations, this decision framework offers a structured roadmap for responsibly leveraging LLMs. Rather than acting as a rigid, one-size-fits-all solution, it shows how advanced language models can be thoughtfully integrated into existing workflows—balancing innovation with accountability to uphold stakeholder trust and regulatory integrity.

\end{abstract}


\section{Introduction and motivation}

The rapid rise of Large Language Models (LLMs) has attracted significant attention across various industries, including finance, where domain-specific language understanding is increasingly essential. These advanced models promise a variety of gains—from more flexible client interactions and streamlined compliance to deeper analytics of complex financial documents. However, these advantages come with multidimensional challenges in data governance, operational costs, fairness, and regulatory risk. In high-stakes contexts such as credit assessment, fraud detection, and investment advisory, financial institutions must not only weigh potential performance benefits but also consider explainability, compliance with sector-specific laws, and the large-scale infrastructure required to maintain reliable LLM-based systems.

 Recent reviews in the financial domain partially address these needs from different angles. For instance, \citet{li_large_2023} and \citet{li_large_2024} present practical surveys highlighting existing LLM solutions in finance and proposing decision frameworks mainly centered on model selection and implementation feasibility, while \citet{lee_large_2025} offers a comprehensive overview of FinLLMs—their historical evolution, evaluation tasks, and open challenges. These works collectively emphasize important aspects of LLMs in finance, such as specialized benchmarking and domain-specific model performance. Meanwhile, more technically oriented reviews discuss parameter-efficient fine-tuning, continual learning, or hallucination-reduction methods \citep{han_parameter-efficient_2024, de_lange_continual_2022, ji_survey_2023}, and another group of policy and ethics papers outlines responsible-AI principles \citep{liu_trustworthy_2024, yang_survey_2023} or focuses on fairness and explainability \citep{guidotti_survey_2018, mehrabi_survey_2022}. Although these studies offer valuable insights into different facets of LLM adoption, they seldom incorporate early feasibility checks—such as determining whether an advanced LLM is truly needed—into the final governance mechanisms and day-to-day oversight for financial institutions.

To close this gap, we propose a structured decision-making framework comprising six interconnected decisions, each vital for responsible LLM adoption in financial contexts. This framework emphasizes a wider lifecycle than is typically considered: from the preliminary question of whether an LLM offers clear benefits over simpler methods, to setting robust data governance policies, to ensuring comprehensive risk management, ethical oversight, Return on Investment (ROI) justification, and deployment strategies. Critically, our approach aims to integrate these concerns so that early-phase decisions—like the level of model complexity or the treatment of sensitive data—directly shape how the institution implements, monitors, and refines its LLM solutions over time.

By merging strategic and technical dimensions, our framework delivers distinct benefits for multiple stakeholders. Executives can align AI initiatives with corporate objectives and resource constraints; compliance teams gain clearer guidance on managing regulatory mandates and data privacy; data scientists learn how to balance advanced modeling with real-world constraints such as interpretability and cost; and ethics committees acquire a proactive tool to tackle bias, transparency, and broader societal impacts. This adaptability makes the framework applicable across a spectrum of organizations, from large global banks with extensive in-house AI capacity to smaller fintech firms or regional institutions needing a more scaled-down approach.

In the sections that follow, we first contextualize LLM usage in finance, outlining practical applications and common pitfalls. Subsequently, each of the six decision areas is explored in detail, clarifying the critical dimensions institutions must consider to successfully implement and manage LLMs. The discussion culminates in a cohesive decision flowchart, visually demonstrating how decisions interconnect and highlighting how iterative feedback loops ensure continuous adaptation and improvement. Ultimately, readers will gain a clear, practical roadmap for responsibly harnessing LLM capabilities—balancing technological innovation with robust compliance, ethical integrity, and sustainable strategic value.
\section{The Opportunities and Risks of LLMs in Finance}
Imagine a dynamic trading floor in a global bank where analysts no longer sift through countless earnings reports but instead employ LLMs to extract insights from expansive disclosures and financial forecasts. This scenario, once speculative, is becoming a practical reality due to advancements rooted in large-scale neural architectures, exemplified initially by GPT-3 \citep{brown_language_2020}. Early on, researchers found that scaling model parameters and training on extensive text corpora induced general-purpose capabilities, often unlocking few-shot mastery of tasks ranging from translation to question answering with minimal textual prompts. These so-called “foundation models” also brought attention to broader risks, including emergent biases and domain-mismatch challenges \citep{bommasani_opportunities_2022}. In finance, domain mismatch is particularly concerning: general-purpose LLMs may struggle on precise terminology, structured tabular data, or the elaborate disclaimers common in regulatory filings \citep{lee_large_2025}. This tension has driven specialized adaptations and opened new research frontiers.

The earliest financial applications drew on BERT \citep{devlin_bert:_2019}, leading to domain-specific variants such as FinBERT \citep{araci_finbert:_2019,yang_finbert:_2020,peng_is_2021,huang_<span_2023}. Instead of rebuilding language encoders from scratch, researchers used domain-specific pretraining on financial news and analyst reports, achieving notable performance gains in sentiment analysis, ESG identification, and risk classification. Subsequent innovations, notably FinGPT \citep{yang_fingpt:_2023}, PIXIU \citep{xie_pixiu:_2023}, and BloombergGPT \citep{wu_bloomberggpt:_2023}, further expanded  coverage to financial data, regulatory filings, and internal transaction records. Concurrently, new bilingual and multimodal initiatives—exemplified by ICE-PIXIU and open-finLLMs projects \citep{hu_no_2024,xie_open-finllms:_2024,xie_finben:_2025}—expanded the capability to handle multiple languages, time-series data, and visual market charts. These advances showed that domain adaptation significantly curtails misclassification and enhances factual grounding, but key challenges persist: numeric reasoning remains difficult, factual “hallucinations” can occur in extended text generation, and certain real-time analyses require continuous updates \citep{chen_finqa:_2022,ji_survey_2023}.

To mitigate the costs and retraining burdens, researchers introduced a variety of parameter-efficient fine-tuning strategies. Low-Rank Adaptation (LoRA) \citep{hu_lora:_2021} appends trainable low-rank matrices to frozen layers, BitFit \citep{zaken_bitfit:_2022} adjusts only bias parameters, and AdapterFusion \citep{pfeiffer_adapterfusion:_2021} chains multiple domain-specific adapters without catastrophic forgetting. Such techniques enable financial institutions to refine large models efficiently, avoiding excessive GPU usage and the necessity to store entire parameter sets for each specialized task. Further optimization avenues appear in knowledge distillation, which compresses huge teacher LLMs into smaller, domain-ready students \citep{sanh_distilbert_2020,gou_knowledge_2021}. Some teams are also exploring retrieval-augmented architectures to keep the model’s knowledge consistent with real-time data, mitigating outdated or incorrect content \citep{lewis_retrieval-augmented_2020,chapman_towards_2022}. Nevertheless, ongoing oversight is paramount. Major policy statements or compliance reports cannot tolerate unverified statements. In banking and insurance, even a single hallucinated risk factor can trigger costly errors \citep{chen_hallucination_2023,ji_survey_2023}.

Beyond technical breakthroughs, using LLMs in finance introduces crucial ethical, regulatory, and fairness challenges. Machine-learning algorithms have faced well-documented critiques for perpetuating discrimination against certain groups \citep{barocas_big_2016}. Within complex financial domains—such as credit underwriting, investment advice, or automated trading—LLMs may unintentionally amplify historical biases present in training datasets \citep{mehrabi_survey_2022,ferrara_should_2023,zheng_judging_2023}. Additionally, \citet{ross_llm_2024} reveal that these models may deviate from purely rational economic principles, reflecting behavioral biases similar to human decision-makers. Such deviations pose challenges for applications requiring strict logical or quantitative rigor. Researchers have proposed multiple mitigation strategies—from re-sampling training data and adversarial testing \citep{wallace_universal_2021} to adopting comprehensive auditing frameworks. These audits occasionally utilize a “model judge” approach, where a secondary LLM identifies inconsistencies or hidden stereotypes within the primary model’s outputs \citep{askell_general_2021,kirk_understanding_2024,mokander_auditing_2024-1}. Concurrently, compliance-focused designs aim to translate complex regulations—like Basel III or General Data Protection Regulation (GDPR)—into actionable code \citep{cao_large_2024}. Practically, regulatory experts note a potential conflict: large generative models might inadvertently learn sensitive or proprietary data, elevating compliance risks if such data is accidentally disclosed \citep{carlini_extracting_2021,rho2024encryption}. Consequently, federated learning and encryption-friendly solutions have gained traction among institutions cautious of centralizing private ledgers or client information \citep{long_federated_2020,che2023federated,yao_federated_2024}.

Alongside fairness and privacy, sustainability concerns loom large. Training or continuously refining domain-specific LLMs consumes vast computational energy, a point quantified by \citet{strubell_energy_2020} and reiterated in cost–benefit analyses that weigh the ROI of LLM usage \citep{shekhar_towards_2024,xexeo_economic_2024,irugalbandara_scaling_2024}. The prospect of regularly re-tuning a 50-billion-parameter model on updated data feeds might be environmentally prohibitive, prompting efforts to exploit smaller or quantized solutions \citep{dettmers_qlora:_2023}. Even so, demand for real-time analytics could intensify. For example, research on anomaly detection in general ledger data shows that LLM embeddings can boost detection of suspicious entries, but they must be regularly updated to stay aligned with changing transaction patterns \citep{bakumenko_advancing_2024}.As organizations weigh capital investment in these AI pipelines, the MLOps frameworks—complete with interpretability, version control, and risk checks—must handle evolving tasks under high transparency \citep{de_lange_continual_2022}.

Amid these multifaceted developments,financial professionals are already experiencing tangible shifts in their daily workflows. Institutions now apply LLMs across diverse tasks—from algorithmic trading and personalized wealth management to automated regulatory compliance, fraud detection, risk assessment, and financial reporting—thereby enhancing automation and human-machine collaboration. Evidence from \citet{lakkaraju_llms_2023} shows that LLM-based advisors can produce inconsistent answers across user profiles, raising questions of fairness in personal finance guidance. Furthermore, \citet{lo_can_2024} underscore the complexities inherent in deploying generative AI for personalized finance, emphasizing the necessity for domain-specific expertise, alignment with client values, and rigorous regulatory compliance. Meanwhile, major banks run pilot programs for summarizing regulatory changes or generating draft earnings commentary \citep{chapman_towards_2022}. Tools like FinGPT \citep{yang_fingpt:_2023} or BloombergGPT \citep{wu_bloomberggpt:_2023} promise advanced capabilities, but they still require thorough domain calibration—especially in dynamic markets. Robust, real-time use cases may require retrieval-based reasoning \citep{lewis_retrieval-augmented_2020,chen_finqa:_2022} combined with meta-learning or continual learning frameworks \citep{finn_model-agnostic_2017,langedijk_meta-learning_2022} to adapt quickly to emerging data. Nevertheless, every innovative method must undergo detailed evaluation by subject matter experts, auditors, and regulatory entities to uphold integrity and ensure compliance with established standards

In summary, the integration of LLMs into financial services is a substantial but manageable undertaking. It emerges from the early finding that scaling up model parameters can yield remarkable versatility \citep{brown_language_2020}, spurring the development of specialized solutions like FinBERT \citep{araci_finbert:_2019}, FinLLMs \citep{li_large_2023}, and more advanced architectures such as GPT-4 \citep{openai_gpt-4_2024}. As these models evolve, the financial sector must carefully adapt its practices around risk management, interpretability, and regulatory compliance \citep{mccarthy_regulation_2022, crisanto_regulating_2024}. While these innovations offer significant potential for improved analytics and automation, they also demand careful oversight and consistent domain alignment, particularly under heightened requirements for data security and fairness

\section{A decision-centric blueprint for LLM implementation in finance}
We introduce a concise, six-decision framework to guide financial institutions through the primary considerations in deploying LLMs. This framework walks practitioners through evaluating whether advanced language capabilities are even warranted, formalizing data governance, establishing essential risk controls, incorporating ethics reviews, determining the proper return on investment, and ultimately selecting and adapting a suitable LLM solution. 

\decisionheading{1}{Rethinking LLMs for simpler solutions}
\label{decision:1}

Before adopting LLMs, financial institutions should carefully evaluate whether these models truly surpass simpler, more transparent methods for a given use case. Although GPT-style architectures and domain-focused BERT variants excel at contextual language interpretation and coherent text generation, many routine financial tasks can still be handled effectively by traditional NLP pipelines or rule-based systems. A core challenge is to weigh LLMs’ advanced linguistic capabilities against their higher computational costs, operational overhead, and opacity—especially when rigorous audit trails and regulatory compliance are nonnegotiable.

LLMs particularly excel in handling unstructured text, for example when analyzing earnings reports, legal documents, or nuanced investor communications. They can detect subtleties in sentiment, implied meanings, or even colloquial expressions that conventional NLP methods might overlook \citep{li_large_2023, xie_pixiu:_2023}. Additionally, their strength as few-shot learners allows them to rapidly generalize from limited examples, making them highly adaptive for novel tasks \citep{brown_language_2020}. Hence, they are especially valuable for interpreting ambiguous user queries, conducting sophisticated market sentiment analyses, or handling intricate compliance documentation \citep{bommasani_opportunities_2022}. Nonetheless, the inherent variability of LLM outputs can be problematic where consistency and deterministic results—like routine compliance checks or structured data extraction—are paramount.

Interpretability also complicates LLM adoption. Given their "black-box" nature, LLMs frequently obscure their reasoning processes, posing significant challenges when justifying their decisions to auditors and regulators \citep{guidotti_survey_2018}. Conversely, traditional NLP or rule-based systems inherently provide clearer, more auditable decision logic, greatly facilitating compliance \citep{rudin_stop_2019}. Institutions must carefully assess whether interpretability is essential for their specific applications, as addressing this through additional interpretability tools or hybrid models inevitably increases complexity and cost.

Another key consideration is resource usage and cost constraints. The computational resources required for training and maintaining LLMs can be substantial, and the associated infrastructure or API costs—particularly under high-volume use—may quickly offset incremental performance gains. In contrast, traditional NLP solutions often deliver sufficiently strong performance with substantially lower overhead. Institutions must rigorously evaluate whether resource investments are justified, particularly when simpler, cost-effective methods might suffice.

Regulatory constraints also play a significant role in determining the appropriateness of LLM adoption. Financial institutions frequently need explicit and verifiable explanations for model-driven decisions, especially in sensitive areas such as credit scoring, risk management, or fraud detection \citep{mccarthy_regulation_2022}. Because LLMs can occasionally produce unpredictable or difficult-to-explain outcomes, firms may need supplementary measures—such as human oversight and robust stress-testing protocols—to comply with regulatory expectations, further increasing costs and operational complexity.

Considering these factors, numerous institutions see a hybrid configuration as the most pragmatic path. By pairing deterministic rules for compliance-critical decisions with LLM-driven language insights, organizations can leverage the best of both approaches. For instance, an LLM might initially evaluate complex client communications for potential compliance risks, while a deterministic layer would finalize the classification or escalation decisions, ensuring auditability and consistency.
\begin{figure}[t!]
  \centering
  \fbox{\includegraphics[width=\textwidth]{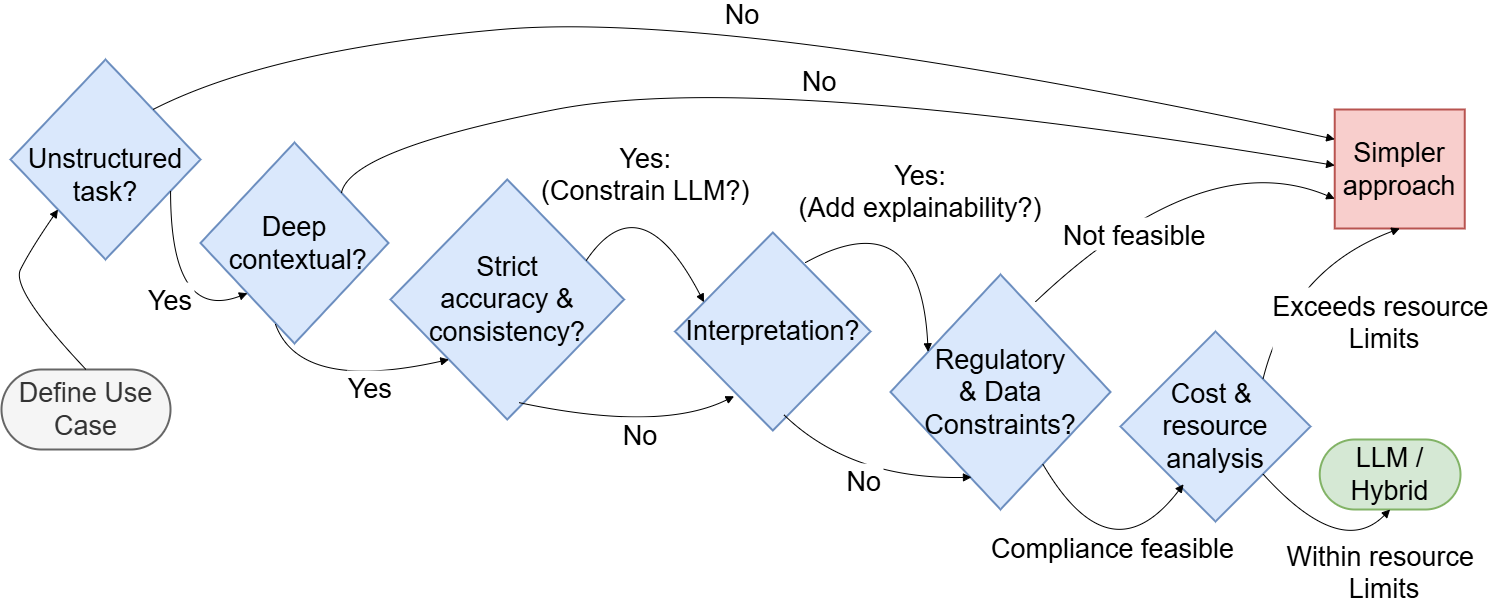}}
  \caption{Decision flowchart for choosing between LLM vs. simpler approaches.}
  \label{fig:decision1_flowchart} 
\end{figure}

The flowchart at Figure \ref{fig:decision1_flowchart} provides a structured pathway for deciding if a LLM is the right tool for a given financial use case. Institutions first assess whether a task involves predominantly unstructured and language-intensive content requiring deep contextual reasoning. If these capabilities are unnecessary, a simpler and more transparent rule-based NLP approach is advisable. For tasks requiring advanced linguistic interpretation, organizations must then rigorously evaluate the necessity for strict accuracy and consistency, the degree of interpretability required by stakeholders and regulators, and the regulatory feasibility of employing an LLM. Finally, the decision process examines resource and cost constraints, ensuring that any incremental gains in linguistic sophistication and analytical capability justify the additional computational resources and financial investments. If these criteria are met, an LLM or a carefully structured hybrid model emerges as a viable solution. Regardless of the chosen approach, ongoing oversight and regular validation remain essential to manage risks like bias, interpretability limitations, and cost escalations effectively.

\decisionheading{2}{Data governance essentials}
\label{decision:2}

Adopting an LLM in financial institutions and fintechs calls for a well-rounded strategy that goes beyond purely technical factors. It demands careful focus on data governance, regulatory obligations, and alignment with broader business goals. Whether a global bank dealing with multifaceted compliance demands or a fintech startup rapidly adopting AI, both must make key decisions about data handling, storage, and regulatory adherence.

Financial institutions often handle highly sensitive customer data, from transaction histories and identity records to confidential corporate communications. Such data is heavily regulated, subject to privacy laws like GDPR \citep{GDPR2016}, banking secrecy provisions, and sector-specific mandates. The first step in responsibly integrating LLMs is rigorous data classification: distinguishing data as public, confidential, or mission-critical. Next, institutions must confirm the legality of letting an LLM process this data, enforcing safeguards such as anonymization, encryption, tokenization, or secure enclaves to minimize unauthorized disclosure. This approach is critical, as research indicates that LLMs can unintentionally memorize and potentially disclose sensitive training data \citep{brown_language_2020, ji_survey_2023}.

Data localization rules also shape how institutions choose their deployment architectures. In numerous jurisdictions, strict rules dictate that customer information must remain within national borders, complicating cross-border transfers. As highlighted by \citet{crisanto_regulating_2024}, financial institutions must carefully assess the legality of even transiently sending sensitive data to external cloud providers or other third-party services. As a result, public-cloud solutions may be unworkable unless robust security measures (e.g., strong encryption and strict access controls) are in place. Conversely, fully on-premises deployments, while providing enhanced data control, impose significant infrastructure and operational costs. Thus, a hybrid arrangement is often ideal: institutions can run high-sensitivity functions (e.g., credit underwriting or wealth management) on-premises or in private clouds, while less critical tasks may use more economical public-cloud services. Techniques like Retrieval-Augmented Generation (RAG), which keep sensitive data local while utilizing cloud-based inference for processing, offer practical solutions to balance innovation with compliance \citep{lewis_retrieval-augmented_2020, lakkaraju_llms_2023}. Even when public-cloud services are acceptable, it is critical that contracts explicitly address security standards, data residency requirements, and business continuity obligations \citep{bcbs2018}. Smaller fintechs may favor external clouds to rapidly deploy advanced AI, but they must watch for vendor lock-in or abrupt policy changes (see \nameref{decision:6}).

Regardless of the chosen deployment architecture, thorough documentation and transparent auditing are indispensable. Financial regulators, such as the European Banking Authority (EBA), require institutions to maintain detailed records of data lineage, model updates, and decision processes \citep{eba2021}. Beyond regulatory requirements, clear, auditable documentation also underpins internal governance and risk oversight (as expanded in \nameref{decision:3}). Methods such as specialized encryption \citep{rho2024encryption} or federated learning \citep{che2023federated, long2020federated} further bolster data security, though they may introduce complexity and higher implementation costs that institutions must consider in their cost-benefit analyses (\nameref{decision:4}).

Effective governance structures unify these concerns by assigning explicit roles and responsibilities—outlining who authorizes data usage, manages audits, and trains staff to prevent accidental leaks. Transparent and ongoing communication with regulators builds essential trust and enables institutions to respond effectively to changing regulatory expectations. As AI regulations shift quickly, sustained monitoring and adaptable deployment choices let financial institutions and fintechs harness LLM benefits without violating privacy laws or internal risk boundaries. Striking this balance calls for careful decisions about model hosting and data-flow governance, establishing a foundation for safe innovation and lasting compliance.

\decisionheading{3}{Risk management for LLMs}\label{decision:3}

Financial institutions adopting LLMs require risk-focused governance structures that address these models’ unique complexity and opacity, complementing earlier data governance strategies. Unlike simpler algorithms, LLMs rely on extremely large parameter sets, complicating efforts to interpret or audit decisions 
 \citep{rudin_stop_2019, barocas_big_2016}. Given the regulatory emphasis on interpretability—particularly in decisions affecting customers and compliance—institutions must carefully balance the powerful capabilities of LLMs with their inherent opacity.

While some researchers recommend using only interpretable models to meet regulatory demands \citep{rudin_stop_2019}, many financial applications require sophisticated language capabilities that are rarely found in completely transparent models. Instead, a more pragmatic solution involves employing explainability frameworks such as Local Interpretable Model-agnostic Explanations (LIME) and Shapley Additive exPlanations (SHAP), or techniques like chain-of-thought prompting, to provide partial transparency into model reasoning \citep{wei_chain--thought_2023}. Emerging approaches, including parameter-efficient fine-tuning and open-source financial LLMs (e.g., FinGPT), allow deeper inspections of specific layers, enabling the creation of valuable accountability artifacts like training records, model adaptations, and detailed prompts \citep{hu_lora:_2021, pfeiffer_adapterhub:_2020, yang_fingpt:_2023}. Integrating these tools into formal governance frameworks ensures that domain experts and compliance officers can confidently verify a model's interpretability level prior to approval and deployment \citep{mccarthy_regulation_2022}.

Traditional Model Risk Management (MRM) practices often struggle with LLM governance, as third-party pretrained models typically provide limited visibility into their internal workings or training data \citep{lee_large_2025}. To address this, institutions are shifting towards adaptive governance strategies that emphasize continuous monitoring and iterative validation post-deployment. This approach includes rigorous pilot testing within sandbox environments, real-time monitoring to detect performance drift, biases, or anomalies, and collaborative oversight involving compliance, data governance, and IT departments \citep{mccarthy_regulation_2022}. Continuous monitoring allows institutions to detect and respond promptly to any performance degradation or unexpected behavior in production environments.

Validation of LLMs in financial contexts demands moving beyond traditional metrics like precision or recall, which often fail to capture nuances such as model hallucinations—outputs that are plausible yet incorrect—particularly under adversarial conditions \citep{ji_survey_2023, wallace_universal_2021}. Institutions should incorporate qualitative, context-specific assessments that evaluate the coherence, reliability, and consistency of LLM outputs, especially in high-stakes use cases like regulatory reporting or client communications. Advanced stress-testing techniques, including adversarial prompt testing and simulations of market volatility, are crucial for assessing LLM robustness. Additionally, maintaining a human-in-the-loop oversight mechanism is essential to periodically audit sensitive tasks, enabling expert judgment to identify and correct subtle errors or ambiguous outputs that automated methods may overlook \citep{bommasani_opportunities_2022}. Institutions must diligently document these validation processes, findings, and corrective actions to ensure transparency and facilitate regulatory oversight.

Addressing fairness and bias proactively is another vital governance responsibility, especially in high-stakes activities such as lending and customer segmentation. Regular bias audits should be integrated into the governance framework to confirm equitable outcomes for protected groups. Such proactive measures mitigate reputational and regulatory risks, aligning closely with ethical and fairness considerations discussed further in \nameref{decision:4A}.

In parallel with robust data governance \nameref{decision:2}, operational and cybersecurity risks require targeted safeguards to prevent data leaks, unauthorized disclosures, and adversarial exploits like prompt injection attacks \citep{carlini_extracting_2021, wallace_universal_2021}. Effective security governance involves stringent access management, comprehensive logging, and robust anomaly detection systems. Implementing advanced methods like federated learning or retrieval-based architectures can significantly reduce data exposure, ensuring compliance with stringent data protection regulations such as GDPR \citep{lewis_retrieval-augmented_2020, long_federated_2020}. Moreover, clear incident-response plans defining escalation pathways, isolation protocols, and rollback strategies further safeguard institutions against potential breaches.

Given shifting regulations and market conditions, dedicated committees must reevaluate LLM performance and compliance expectations at regular intervals, updating risk controls as needed. Comprehensive documentation of all governance activities, including risk evaluations, approvals, and corrective actions, provides transparency and accountability essential for both regulatory scrutiny and internal stakeholder confidence. Table ~\ref{table:Dec3} outlines key governance areas, clearly highlighting the importance of each factor, identifying essential questions for practitioners, recommending effective mitigation steps, and pinpointing potential issues if these measures are not addressed. This approach helps institutions proactively adapt to changing regulatory expectations and stakeholder needs.

\begin{table}[ht]
\centering
\caption{Key governance dimensions for responsible LLM}
\footnotesize
\setlength\tabcolsep{3pt}
\begin{tabular}{p{2.4cm} p{2.3cm} p{3.2cm} p{4.7cm} p{2.8cm}}
\toprule
\textbf{Dimension} & \textbf{Why It\newline Matters} & \textbf{Key Questions} & \textbf{Approaches} & \textbf{Failures} \\
\midrule

\textbf{Interpretability} 
& High-stakes usage, regulatory demand for rationale 
& Are outputs explainable for audits?\newline 
Partial or chain-of-thought reasoning? 
& Gated MRM sign-offs\newline 
Local explainer tools (e.g.\ SHAP)\newline 
Adapter-based or open-source 
& Opaque logic\newline 
Noncompliance with transparency 
\\
\hline

\textbf{Validation} 
& Risk of hallucinations, adversarial inputs 
& How to ensure domain accuracy?\newline 
Any stress tests for vulnerabilities? 
& Domain benchmarks, red-team prompts\newline 
Human-in-the-loop checks\newline 
Detailed validation logs 
& Missed logic flaws\newline 
Gaps in traceability 
\\
\hline

\textbf{Bias Audits}
& Fair decisions, reputational harm 
& Are bias audits part of governance?\newline 
Any adverse impact on protected groups? 
& Formal bias checks\newline 
Fine-tuning for disparate impact\newline 
Audit logs for compliance 
& Legal violations\newline 
Erosion of trust 
\\
\hline

\textbf{Security} 
& Data leakage, prompt injection 
& Is data minimized/protected?\newline 
Incident-response plan in place? 
& Role-based controls, logging, detection\newline 
On-prem or federated approaches\newline 
Encryption, data minimization 
& Breaches of sensitive info\newline 
Delayed malicious-response handling 
\\
\hline
\textbf{Adaptive Lifecycle Governance} 
& LLMs can exhibit drift, require iterative refinements 
& Are pilot/sandbox tests used before full deployment?\newline How to handle continuous updates or new regulations? 
& Controlled pilot phases;\newline 
Continuous post-deployment monitoring;\newline 
Frequent re-checks with domain experts;\newline
Iterative governance committees 
& Outdated or stale model outputs;\newline
Unnoticed performance drop;\newline
Late response to new laws 
\\
\hline
\textbf{Ongoing Oversight} 
& Evolving rules, lifecycle adjustments 
& Who is model owner?\newline 
When to retrain or roll back? 
& Governance committees under MRM\newline 
Scheduled audits for performance\newline 
Documentation of approvals 
& Undetected drift\newline 
Regulatory misalignment 
\\

\bottomrule
\end{tabular}
\label{table:Dec3}
\end{table}

\decisionheading{4}{Ethical oversight for LLM integration}\label{decision:4A}

While \nameref{decision:3} addresses core risk controls, LLM adoption also raises unique ethical questions, such as fairness and long-term societal impact, that transcend routine compliance or operational concerns. Given the unique role of financial institutions as trusted custodians in society, ethical considerations must be placed at the forefront of LLM adoption. Accordingly, organizations must embed formal ethical oversight processes beyond basic compliance, to protect institutional reputation and ensure equitable financial services.

Bias mitigation represents one of the most pressing ethical issues. LLMs trained on historical data inherently risk replicating past inequities—such as systematically disadvantaging particular demographic groups in credit or investment recommendations. Effective bias mitigation demands rigorous pre-deployment testing, applying scenario-based analyses across diverse demographic factors like race, gender, and ethnicity \citep{ mehrabi_survey_2022}. Practical interventions include fairness-driven data augmentation, algorithmic adjustments to equalize model outcomes across demographic groups, and structured human oversight for critical decisions \citep{liang_towards_2021, mokander_auditing_2024-1}. Institutions should adopt established ethical frameworks such as the Monetary Authority of Singapore’s FEAT principles (Fairness, Ethics, Accountability, Transparency), reinforcing bias audits and inclusive development processes.

Building on the data protections outlined in \nameref{decision:2}, ethical oversight adds another layer of accountability, requiring that anonymization and encryption protocols explicitly consider user autonomy.
Transparent communication with customers, clearly outlining how their data is processed by AI systems, combined with robust consent mechanisms, reinforces privacy protections and user trust.

Accountability and transparency remain paramount due to the inherent opacity associated with LLM decision-making. Institutions must clearly define human accountability structures, ensuring a designated individual or team retains ultimate responsibility for AI-assisted decisions \citep{rudin_stop_2019}. Detailed documentation practices—including audit logs, comprehensive model cards describing training datasets, known biases, and model limitations—provide necessary transparency and enable meaningful oversight. Furthermore, transparent disclosures to customers about AI involvement and explicit error-handling procedures enhance public trust and maintain regulatory alignment.

\begin{table}[H]
\centering
\footnotesize
\caption{Key ethical and societal dimensions for LLM adoption }
\label{tab:ethics_summarized_streamlined}
\begin{tabular}{@{}p{0.15\linewidth}p{0.25\linewidth}p{0.25\linewidth}p{0.28\linewidth}@{}}
\toprule
\textbf{Dimension} & \textbf{Key Risk} & \textbf{Adverse Outcome} & \textbf{Mitigation} \\
\midrule

\textbf{Bias \& Fairness} 
& Historical data biases replicated in AI outputs 
& Inequitable credit or investment decisions; reputational harm 
& Pre-deployment bias testing; scenario-based analysis; fairness data augmentation; FEAT adoption \\

\cmidrule(lr){1-4}
\textbf{Privacy \& Data Protection} 
& Mishandling of sensitive personal inforkimation 
& Regulatory breaches (GDPR, CCPA); loss of customer trust 
& Data anonymization; strict access controls; end-to-end encryption; transparent consent processes \\

\cmidrule(lr){1-4}
\textbf{Accountability \& Transparency} 
& Opaque decision-making; unclear human responsibility 
& Limited recourse for AI errors; diminished public confidence 
& Designate accountable roles; keep audit logs/model cards; disclose AI usage; error-handling protocols \\

\cmidrule(lr){1-4}
\textbf{Workforce Transformations} 
& Unplanned automation displacing employees 
& Employee morale issues; skill gaps; labor disputes 
& “AI augmentation” mindset; upskilling/reskilling; clear transition support; open communication \\

\cmidrule(lr){1-4}
\textbf{Customer Autonomy} 
& Overreliance on AI; lack of human override 
& Potential manipulation or confusion; reduced agency 
& Keep AI advisory; allow human intervention; routine testing for manipulative behaviors \\

\cmidrule(lr){1-4}
\textbf{Continuous Governance \& Ethical Oversight} 
& Unchecked model drift; emerging ethical gaps 
& Accumulating biases; slow response to new regulations; reputational damage 
& Ethical committees; iterative impact assessments; stakeholder engagement; periodic re-audits \\

\bottomrule
\end{tabular}
\end{table}
The introduction of LLMs inevitably transforms workforce roles, necessitating proactive ethical management of job impacts. Moreover, oversight committees should monitor not only model performance but also workforce transitions, ensuring staff have the training to supervise AI outputs ethically and handle shifts in job responsibilities. Investment in upskilling and reskilling programs, proactive communication about AI-driven changes, and clear transition support pathways are essential to ethically manage these workforce dynamics, ensuring employees remain engaged, informed, and well-equipped to leverage AI advancements effectively.

Protecting customer autonomy and maintaining trust are critical ethical priorities. Customers should clearly understand when they interact with an AI-driven system and retain the choice to request human intervention. AI-generated financial recommendations must remain advisory rather than authoritative, preserving user agency. Ethical oversight panels should regularly probe for manipulative or deceptive behaviors, ensuring AI suggestions remain transparent and respect customer autonomy.

Finally, continuous governance and ethical oversight underpin responsible LLM adoption. In addition to internal compliance, institutions should establish dedicated ethical committees responsible for regularly assessing AI deployments against established ethical standards and evolving societal expectations \citep{bommasani_opportunities_2022}. As summarized in Table~\ref{tab:ethics_summarized_streamlined}, these dimensions collectively ensure that ethical considerations are woven into every stage of LLM usage. Regular stakeholder engagements, iterative ethical impact assessments, and periodic re-evaluations of bias, privacy, accountability, and autonomy safeguards ensure governance frameworks adapt dynamically to new risks and regulatory landscapes. By embedding ethical oversight as an ongoing process rather than a static checkpoint, financial institutions can responsibly leverage LLM innovations, reinforcing their role as trusted and ethical stewards of advanced technology in finance.

\decisionheading{5}{ Measuring LLM value and ROI}\label{decision:4}

Building on both technical and regulatory foundations, organizations must also examine the broader strategic and financial impacts of LLM deployment. Implementing an LLM is fundamentally a long-term business decision, requiring meticulous cost-benefit analysis to balance immediate efficiency gains and direct financial returns against long-term broader organizational advantages, including brand enhancement, innovation potential, and strengthened customer relationships. Executives must thoroughly justify and align each proposed use case to overall corporate goals, considering both immediate tangible outcomes and intangible benefits that mature over time. Preparing staff to effectively integrate and utilize LLM capabilities from the outset remains equally vital.

Tangible financial returns from LLM adoption frequently stem from efficiency gains or revenue enhancements. Automation of routine processes such as customer support interactions or report generation can substantially decrease operational expenses, from labor costs per transaction to broad process automation. Similarly, LLM-driven product recommender systems can directly generate incremental sales, demonstrating clear financial benefits \citep{chang_survey_2024}. However, these deployments still require careful management of upfront and ongoing expenses, including infrastructure investments, model training, API usage fees, and engineering resources needed for integration with legacy systems. Additional hidden expenses involve extensive data preparation and robust governance frameworks to mitigate risks like model hallucinations, privacy infringements, and cybersecurity vulnerabilities, all of which can negatively impact trust and compliance \citep{shekhar_towards_2024}.

Beyond these measurable impacts, institutions should also assess intangible and strategic dimensions, for example, brand perception, user confidence, and innovation posture, that can significantly influence market positioning. LLMs consistently delivering reliable, personalized, and accurate customer interactions can elevate customer satisfaction scores, foster brand loyalty, and build deeper trust—benefits harder to quantify immediately yet impactful for long-term market positioning \citep{gupta_ai_2024}. Additionally, adopting innovative AI solutions enhances institutional reputation as a technology leader, potentially attracting tech-savvy customers, strategic partnerships, and talented staff. Such intangible benefits, although less immediately quantifiable, profoundly influence sustainable market differentiation. Metrics like Net Promoter Score (NPS), customer effort scores, reputational indices, and perceived reliability help organizations quantify and strategically incorporate these intangible assets  (see Table~\ref{tab:kpi_matrix}).

Given the inherent complexity in accurately forecasting these benefits, institutions should adopt rigorous project management methodologies, such as iterative pilot testing or specialized MLOps lifecycles, to continuously validate ROI against clearly defined Key Performance Indicators (KPIs). This disciplined approach involves establishing baseline measurements prior to deployment and regularly reviewing indicators such as incremental revenue, process automation rate, and customer satisfaction. Initial ROI predictions must be iteratively validated through real-world deployments to ensure ongoing alignment with organizational objectives \citep{shekhar_towards_2024}.

Selecting the appropriate LLM is not solely about choosing the most powerful model but strategically aligning model capabilities with targeted business outcomes. While highly advanced models offer impressive capabilities, they may impose disproportionate costs or complexity relative to specialized, domain-tailored alternatives. Institutions should carefully evaluate whether advanced functionalities justify additional expenses by aligning model choices with clearly defined business objectives and economic returns \citep{xexeo_economic_2024}. Moderately sized models might deliver superior cost-benefit ratios for structured or routine workflows, whereas more sophisticated models might be strategically justified when differentiation and innovation leadership are central priorities.

\begin{table}[t!]
\centering
\small
\caption{Illustrative KPI matrix for LLM ROI and strategic alignment }
\label{tab:kpi_matrix}
\begin{adjustbox}{width=\columnwidth,center}
\begin{tabular}{@{}lll@{}} 
\toprule
\textbf{Dimension} & \textbf{KPI} & \textbf{Description} \\
\midrule

\multirow{4}{*}{\shortstack{\textbf{Economic}}} 
& Process Automation Rate & \% of tasks handled by LLM; labor savings. \\
& Cost per Transaction & Cost/task pre- vs. post-LLM; expense change. \\
& Incremental Sales & Added revenue from LLM recommendations. \\
& ROI & Net gain over investment; ratio $>1$ suggests profit \\

\midrule
\multirow{2}{*}{\shortstack{\textbf{User-Centric}\\(Tangible)}} 
& Task Completion Rate & \% of tasks fully resolved by LLM; effectiveness. \\
& Task Completion Time & Time to finish tasks \\

\midrule
\multirow{4}{*}{\shortstack{\textbf{User-Centric}\\(Intangible)}} 
& NPS & User recommendation likelihood; satisfaction. \\
& Customer Effort Score & Ease of task completion; lower = better. \\
& Reputational Index & Public sentiment on AI; signals trust. \\
& Perceived Reliability & User confidence in LLM consistency/accuracy. \\

\midrule
\multirow{2}{*}{\shortstack{\textbf{Strategic}}} 
& Time-to-Market & Speed of AI feature launch; agility. \\
& AI Adoption Index & \% of units using LLM; cultural integration. \\

\bottomrule
\end{tabular}
\end{adjustbox}
\end{table}

A phased adoption pathway allows institutions to rigorously validate each step and scale thoughtfully. Initial pilot projects with targeted KPIs clarify tangible benefits, informing clear go/no-go decisions. Successful pilots justify broader integration, allowing ongoing alignment with strategic objectives, continuously validated by multidisciplinary teams comprising finance, IT, compliance, and business strategy. Continuous monitoring ensures stable ROI and adaptability to evolving market and regulatory landscapes.

Ultimately, the successful adoption of LLMs emerges through a structured approach that balances immediate gains with sustained strategic impact, supported by disciplined measurement, iterative validation, and adaptive governance. By embedding LLM investments within a comprehensive long-term framework, financial institutions can track ROI through the Key Performance Indicators (Table~\ref{tab:kpi_matrix}), refining both usage and governance processes in tandem.

\noindent

\decisionheading{6}{Choosing the optimal LLM implementation path}\label{decision:6}
Having carefully evaluated strategic considerations, institutions must now thoughtfully select their LLM implementation approach, balancing financial viability, regulatory adherence, and desired customization to best meet their unique operational and sector-specific requirements. A misstep in this process can lead to overspending, vendor dependency, or regulatory issues, highlighting the importance of a thoughtful selection approach. 

\subsubsection*{Open-Source vs. proprietary models:}

Open-source models (e.g. Meta’s LLaMA or financial-domain models like FinBERT or FinGPT)  provide institutions full access to the model’s source code and internal parameters, enabling significant customization and fine-tuning on proprietary data without external exposure risks. This transparency is particularly appealing for banks and fintech firms facing rigorous regulatory scrutiny, as internal teams can thoroughly audit model behavior and directly manage data security. An illustrative case is ANZ Bank, which shifted from using OpenAI’s API to fine-tuning an open-source LLaMA variant, primarily driven by cost management and regulatory concerns. Additionally, open-source models generally offer reduced ongoing licensing costs and benefit from continuous enhancements contributed by the broader community \citep{xie_open-finllms:_2024}. However, they often lag behind state-of-the-art proprietary models in raw performance, demanding substantial internal expertise and resources since external vendor support is unavailable for maintenance and debugging \citep{li_large_2024}.

Proprietary models (such as OpenAI’s GPT-4, Google’s PaLM, or industry-specific ones like BloombergGPT) typically arrive pre-trained and readily accessible through APIs or licensed software. These solutions are particularly appealing to financial organizations seeking swift deployment, ease of maintenance, and minimal technical complexity. They typically deliver immediate high performance, consistent updates, and potentially advanced, specialized architectures beneficial for precise financial tasks. For example, BloombergGPT, trained on a comprehensive financial dataset, demonstrates significant performance advantages on financial benchmarks but involves extensive resource investment inaccessible to most institutions \citep{wu_bloomberggpt:_2023}. 

However, reliance on proprietary solutions introduces distinct challenges. Institutions typically have limited visibility into proprietary models, restricting deep customization. Current API agreements—such as OpenAI’s, which, as of now, prohibit extensive fine-tuning of their largest GPT-4 models—force institutions to rely primarily on prompt engineering or minimal adjustments. This can limit performance on specialized financial tasks. Additionally, dependence on a single vendor exposes financial institutions to risks such as unexpected price hikes, sudden policy shifts, or service disruptions, potentially threatening operational continuity \citep{irugalbandara_scaling_2024}. Sending sensitive data to third-party providers also raises significant compliance and data security concerns, especially regarding regulations like GDPR. While some proprietary vendors offer mitigations like on-premises deployment or data isolation, these options aren't universally available and typically involve higher costs.

Thus, choosing between open-source and proprietary LLMs depends significantly on institutional priorities: Is maintaining full control over data, transparency, and customizability paramount, or does rapid deployment with vendor-managed convenience provide greater strategic value? This choice often evolves as institutions mature, with fintech startups frequently beginning with proprietary models for ease of deployment before transitioning to open-source alternatives as internal AI expertise and infrastructure expand.

\subsubsection*{In-House vs. vendor deployment:} Another important dimension for institutions to consider is whether to manage LLM deployment internally or engage external vendors. This decision, although related, is distinct from the choice between open-source and proprietary models. For instance, an institution might license a proprietary LLM but host and operate it entirely on-premises, or conversely, deploy an open-source model managed entirely through a third-party service provider. Thus, the distinction between in-house and vendor deployment primarily addresses responsibility for implementation, ongoing management, and infrastructure rather than the openness or ownership of the model itself.

Institutions choosing for in-house deployment typically entrust their internal teams with the full scope of model training, integration, and performance management. This strategy grants comprehensive control over the AI lifecycle, particularly advantageous for adhering to stringent regulatory standards like data residency and auditing requirements \citep{crisanto_regulating_2024}. Nevertheless, maintaining such complete oversight comes with substantial resource implications, demanding considerable technical expertise, infrastructure investments, and continuous model validation efforts. Consequently, the institution directly assumes all responsibilities related to model performance, including potential biases and compliance issues. Given these demands, in-house deployment is often most suitable for larger organizations possessing the requisite expertise and financial resources, whereas smaller institutions may find these commitments difficult to sustain.

In contrast, partnering with external vendors often enables financial institutions to leverage pre-trained models and specialized vendor expertise for a faster time-to-market. For instance, Morgan Stanley’s collaboration with OpenAI to integrate GPT-4 into its wealth management platform exemplifies how institutions can outsource much of the model development complexity while benefiting from cutting-edge vendor capabilities. However, vendor dependence raises concerns about service reliability, data security, and operational resilience, issues that regulators consistently underscore \citep{crisanto_regulating_2024}. Although contractual protections (e.g., clear performance benchmarks, data protection standards, compliance auditing clauses) can partially mitigate these risks, ultimate accountability for client outcomes and adherence to regulations remains with the financial institution. Consequently, careful vendor agreement negotiation is crucial to ensure robust data safeguards, transparent service commitments, and clear dispute resolution pathways that align with both institutional needs and regulatory expectations.

In practice, many financial institutions prefer a hybrid implementation strategy, embedding vendor-supplied LLMs within customized internal environments that incorporate encryption and zero-retention mechanisms to strengthen data governance. Concurrently, these institutions often deploy additional models on-premises or blend open-source with proprietary solutions, tailoring their choices according to organizational risk tolerance. This approach provides a practical pathway to accelerated innovation while enabling rigorous oversight of sensitive workflows—a critical consideration for institutions navigating rapid AI adoption within stringent regulatory contexts. Clearly specifying the interactions between external models and internal proprietary data further mitigates integration uncertainties and enhances data security. Over time, especially mid-sized institutions facing dynamic regulatory demands or striving for competitive differentiation through specialized AI services may strategically transition from initial vendor dependence toward fully autonomous in-house AI capabilities.

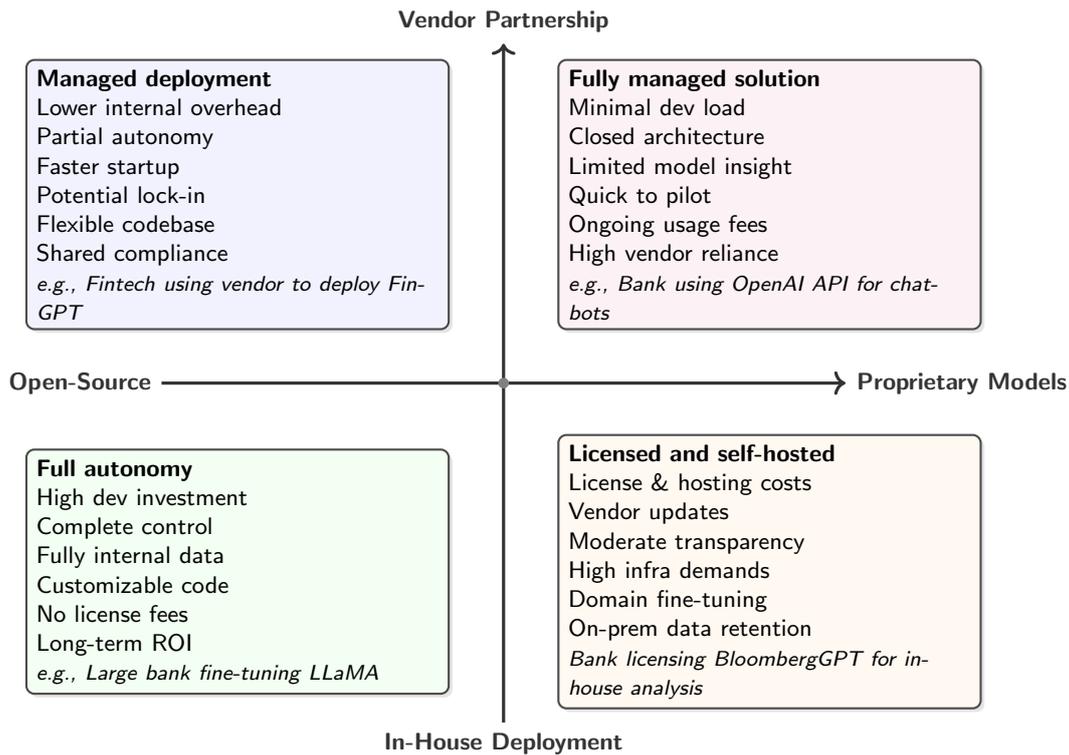
\begin{figure}[ht]
\centering
\begin{tikzpicture}[scale=1.0]
  \draw[->,very thick, black!80] (-4.5,0) -- (4.5,0) node[right,font=\sffamily\bfseries\footnotesize]{\textbf{Proprietary Models}};
  \draw[->,very thick, black!80] (0,-4.5) -- (0,4.5) node[above,font=\sffamily\bfseries\footnotesize]{\textbf{Vendor Partnership}};
  
  \node[font=\sffamily\bfseries\footnotesize, text=black!80] at (-4.5,0) [left]{\textbf{Open-Source}};
  \node[font=\sffamily\bfseries\footnotesize, text=black!80] at (0,-4.5) [below]{\textbf{In-House Deployment}};
  
  \node[
    rectangle,
    rounded corners=3pt,
    draw=black!70,
    fill=blue!5,
    thick,
    text width=5.3cm,
    font=\footnotesize\sffamily,
    align=left,
    drop shadow={shadow xshift=0.5mm, shadow yshift=-0.5mm, opacity=0.2}
  ]
  at (-3.5,2.5)
  {
    \textbf{Managed deployment}\\
    Lower internal overhead\\
    Partial autonomy\\
    Faster startup\\
    Potential lock-in\\
    Flexible codebase\\
    Shared compliance\\
    \textit{\scriptsize e.g., Fintech using vendor to deploy FinGPT}
  };
  
  \node[
    rectangle,
    rounded corners=3pt,
    draw=black!70,
    fill=purple!5,
    thick,
    text width=5.3cm,
    font=\footnotesize\sffamily,
    align=left,
    drop shadow={shadow xshift=0.5mm, shadow yshift=-0.5mm, opacity=0.2}
  ]
  at (3.5,2.5)
  {
    \textbf{Fully managed solution}\\
    Minimal dev load\\
    Closed architecture\\
    Limited model insight\\
    Quick to pilot\\
    Ongoing usage fees\\
    High vendor reliance\\
    \textit{\scriptsize e.g., Bank using OpenAI API for chatbots}
  };
  
  \node[
    rectangle,
    rounded corners=3pt,
    draw=black!70,
    fill=green!5,
    thick,
    text width=5.3cm,
    font=\footnotesize\sffamily,
    align=left,
    drop shadow={shadow xshift=0.5mm, shadow yshift=-0.5mm, opacity=0.2}
  ]
  at (-3.5,-2.5)
  {
    \textbf{Full autonomy}\\
    High dev investment\\
    Complete control\\
    Fully internal data\\
    Customizable code\\
    No license fees\\
    Long-term ROI\\
    \textit{\scriptsize e.g., Large bank fine-tuning LLaMA}
  };
  
  \node[
    rectangle,
    rounded corners=3pt,
    draw=black!70,
    fill=orange!5,
    thick,
    text width=5.3cm,
    font=\footnotesize\sffamily,
    align=left,
    drop shadow={shadow xshift=0.5mm, shadow yshift=-0.5mm, opacity=0.2}
  ]
  at (3.5,-2.5)
  {
    \textbf{Licensed and self-hosted}\\
    License \& hosting costs\\
    Vendor updates\\
    Moderate transparency\\
    High infra demands\\
    Domain fine-tuning\\
    On-prem data retention\\
    \textit{\scriptsize Bank licensing BloombergGPT for in-house analysis}
  };
  
  \fill[black!50] (0,0) circle (2pt);
\end{tikzpicture}
\caption{LLM Implementation Matrix - Balancing Model and Deployment Strategic Options}
\label{fig:implementation_tool}
\end{figure}

Figure \ref{fig:implementation_tool} illustrates how the choice of model openness (horizontal axis) intersects with deployment strategy (vertical axis). Moving horizontally from left to right represents a shift from open-source toward proprietary architectures, while vertically from bottom to top indicates a progression from in-house to outsourced operational management. Each quadrant highlights unique trade-offs regarding control, cost, and regulatory compliance, enabling financial institutions to identify the optimal configuration aligned with their priorities.

\subsection*{Adaptation Methods}  Having identified whether an open- or closed-source model will be used—and whether it will be hosted in-house or via a vendor—institutions must then determine the most suitable adaptation strategy for their specific tasks. In practice, however, these two decisions are closely interwoven: certain hosting choices may limit or enable particular adaptation methods, and vice versa. Nevertheless, for clarity of exposition, we outline the main adaptation pathways here, ranging from simply using a pretrained model “as-is” to full retraining on domain data. Each option brings different trade-offs in technical complexity, cost, and performance, requiring organizations to balance the benefits of deeper customization against the constraints of their chosen deployment model.

\textbf{Prompt engineering} involves carefully designing input prompts to guide an LLM’s outputs without modifying its internal parameters. By precisely specifying instructions or embedding relevant examples within these prompts, practitioners effectively leverage the existing knowledge of the model, circumventing the complexity of parameter retraining. Research into in-context learning illustrates that even a few well-chosen examples within a prompt can enable LLMs to swiftly adapt to new or highly specialized tasks \citep{brown_language_2020}. This flexibility supports rapid iteration, allowing quick adjustments as task requirements evolve.

In financial applications, prompt engineering stands out for its agility and low implementation risk. Financial institutions can quickly pivot between tasks—such as summarizing regulatory updates on one day and addressing client portfolio queries the next—by simply modifying the prompts. This approach is cost-effective, requiring only basic inference-time resources and minimal infrastructure, significantly simplifying data governance and operational management. However, there is an often-overlooked investment in the staff time and expertise required to iteratively refine prompts, particularly to ensure outputs consistently reflect precise financial terminology or regulatory compliance disclosures.

Despite its benefits, prompt engineering does have inherent limitations. If the underlying LLM lacks specialized financial knowledge, even carefully crafted prompts may yield inconsistent or incomplete responses, posing notable risks in critical contexts such as fraud detection or compliance reporting. Moreover, financial institutions leveraging third-party LLM APIs must carefully manage prompts containing sensitive or private financial information, adhering strictly to data protection standards and regulatory requirements. Therefore, institutions must implement rigorous internal procedures to consistently verify the compliance and accuracy of prompt-engineered outputs. 

\textbf{Full fine-tuning} involves retraining all parameters of a pretrained LLM using task- or domain-specific datasets, such as proprietary financial documentation, specialized Q\&A pairs, or compliance manuals \citep{araci_finbert:_2019,wu_bloomberggpt:_2023}. This process embeds precise domain knowledge into the model itself, allowing financial institutions to achieve higher accuracy and contextual richness for sophisticated tasks like credit-risk modeling, underwriting decisions, or complex advisory services.

A primary advantage of full fine-tuning lies in its ability to deeply internalize specialized financial jargon, support intricate analyses, and consistently adhere to rigorous compliance standards—capabilities often beyond general-purpose models. For example, tasks like crafting detailed investment recommendations or automatically incorporating region-specific regulatory disclaimers become significantly more reliable through fully fine-tuned variants. However, achieving these benefits demands substantial investments, including powerful GPU infrastructure, sophisticated data pipelines, and robust compliance monitoring frameworks. Additionally, each new fine-tuned model typically requires dedicated storage and regular audits, a particularly critical consideration given the dynamic nature of financial markets, where regulatory shifts or new product introductions frequently necessitate model retraining.

From a practical standpoint, smaller firms may opt to delegate fine-tuning activities to external providers, but this approach raises concerns around data privacy, dependency on vendor-provided updates, and reduced control over sensitive datasets. Nevertheless, when high accuracy and precise, domain-specific reasoning are essential—for instance, in performing sophisticated risk assessments or generating legally compliant financial documents—the strategic advantages of full fine-tuning often justify the associated costs. Ultimately, although embedding specialized expertise directly into the model enables optimization of critical workflows, it also imposes ongoing maintenance demands and reduces flexibility to adapt quickly to new tasks without additional training.

\textbf{Parameter-Efficient Fine-Tuning (PEFT)} adapts a LLM by selectively training only a small subset of parameters, while most layers remain frozen. Techniques such as LoRA \citep{hu_lora:_2021}, adapters \citep{houlsby_parameter-efficient_2019}, and prefix tuning \citep{li_prefix-tuning:_2021} introduce lightweight modules or modify select weights to infuse domain-specific knowledge efficiently, significantly reducing computational requirements \citep{han_parameter-efficient_2024}.

In financial contexts, PEFT effectively captures specialized content without the extensive computational overhead associated with full fine-tuning. The core model maintains broad linguistic and general reasoning capabilities, while the added modules handle nuanced, domain-specific tasks. Although maximum achievable accuracy might be slightly lower compared to fully fine-tuned models, the substantial reduction in training costs and streamlined storage needs often makes this compromise acceptable for many financial institutions.

Cost-wise, PEFT is notably economical: adapter modules or low-rank matrices typically require minimal storage space, are faster to train, and demand fewer GPU resources. This affordability particularly benefits smaller organizations or projects with constrained AI budgets. Nevertheless, adopting PEFT necessitates access to a base model that permits modifications—usually an open-source or internally hosted solution—since proprietary API providers typically restrict alterations to model weights.

Compliance and regulatory considerations remain critical, even with partial adaptations. Institutions must thoroughly validate and audit each modified module, especially when different modules handle sensitive workflows such as retail banking versus institutional trading. Moreover, the modular nature of PEFT simplifies incremental updates, enabling rapid adjustments to changing regulatory environments or market demands without needing extensive retraining.

In summary, parameter-efficient tuning offers a balanced approach between prompt engineering and full fine-tuning. Within finance, it supports agile and frequent adaptations driven by regulatory or market shifts while avoiding the significant computational and governance burdens associated with retraining entire models from scratch.

\textbf{Retrieval-Augmented Generation (RAG)} combines a LLM with an external knowledge retrieval system \citep{lewis_retrieval-augmented_2020}. Rather than relying exclusively on the model’s internal parameters, the LLM can fetch relevant documents from a separate repository—such as an internal research database, policy collection, or curated news feed—when generating responses. Typically, this involves embedding the user’s query, locating the nearest matching documents through a vector-based index, and prepending those texts to the model’s prompt. By referencing this short-term “memory,” the system integrates up-to-date and domain-specific information, greatly enhancing factual accuracy and reducing reliance on the model’s potentially outdated internal weights.

In the financial sector, RAG’s main advantage is the ability to inject real-time knowledge on demand, making it ideal for tasks like responding to new regulatory changes or analyzing current market data. Because the model consults authoritative documents rather than relying solely on training data, factual precision improves and the likelihood of hallucinations decreases. A single LLM can thus support multiple applications—from compliance checks to product inquiries—as long as relevant documents are available. However, the institution must maintain a robust knowledge base and index: if certain references are incomplete or outdated, the model’s answers will suffer. Additionally, the need to manage longer prompts and carefully merge retrieved snippets imposes extra engineering overhead to preserve output coherence.

On the cost side, RAG alleviates the burden of continuously retraining a large model by keeping new knowledge external, thereby avoiding repeated updates to model weights. Although building and maintaining a reliable retrieval pipeline requires ongoing collaboration between subject matter experts and engineering teams, overall expenses remain lower than regularly fine-tuning the entire LLM whenever data changes.

From a regulatory perspective, RAG offers clear traceability. Each answer can be tied back to a specific source document, simplifying auditing and clarifying how conclusions were formed. Moreover, since no newly ingested data is permanently “baked” into the model, the risk of unintentionally memorizing private information decreases. Even so, any snippets passed to an externally hosted LLM must be scrutinized for confidentiality, especially if they include sensitive client data—requiring additional safeguards around user permissions and data encryption. Finally, institutions must validate not only the language model but also the quality and completeness of the retrieval system, given that retrieval gaps or misleading documents can propagate erroneous answers to critical financial queries.

\begin{table}[t!]
\centering
\footnotesize
\caption{Comparison of LLM adaptation methods}
\label{tab:adaptation_methods_refined}
\setlength{\tabcolsep}{3pt}
\begin{tabular}{@{}p{2.8cm} p{3cm} p{3cm} p{3cm} p{3cm}@{}}
\toprule
\textbf{Dimension} & \textbf{Prompting\newline (No Tuning)} & \textbf{PEFT} & \textbf{Full Fine-Tuning} & \textbf{RAG}\\
\midrule

\textbf{Core Idea}
& Use only prompts; no parameter updates
& Train small add-on modules; freeze rest
& Retrain all model weights on domain data
& Fetch external documents in real time\\
\midrule

\textbf{Compute \& Cost}
& Minimal; inference-only
& Moderate; smaller modules
& High; expensive retraining + storage
& Low--moderate; maintain search index\\
\midrule

\textbf{Performance}
& Limited by base model/prompt quality
& Often near full-fine-tune accuracy
& Potentially highest if data are robust
& Strong, given up-to-date external sources\\
\midrule

\textbf{Implementation Speed}
& Rapid prompt design
& Faster updates than full tuning
& Slower: large-scale retraining
& Moderate: set up retrieval system\\
\midrule

\textbf{Hallucination Risk}
& Relatively high if prompts unclear
& Reduced drift via domain modules
& Generally lower if domain-trained
& Lowest: references to authoritative docs\\
\midrule

\textbf{Explainability}
& Low; opaque base model
& Medium; adapters can be inspected
& Low--medium; knowledge embedded in weights
& Higher; references clarify sources\\
\midrule

\textbf{Data \& Privacy}
& No new data in model; prompts may leak info
& Domain info in small modules
& Full model may store sensitive data
& Sensitive data external; retrieved as needed\\
\midrule

\textbf{Best Fit}
& Rapid prototyping; broad tasks
& Cost-effective customization
& Mission-critical, domain-specific tasks
& Dynamic or compliance-heavy scenarios\\

\bottomrule
\end{tabular}
\end{table}

As shown in Table \ref{tab:adaptation_methods_refined}, the most suitable LLM adaptation strategy depends on each institution’s data requirements, budget constraints, and compliance obligations. Crucially, these methods are not mutually exclusive. A financial firm with rapidly changing demands might deploy RAG for up-to-date references while using parameter-efficient tuning to embed specialized compliance rules and applying prompt engineering to fine-tune instructions on short notice. Where maximum control over outputs is required—such as in regulated advisory or automated underwriting—full fine-tuning can be introduced to further integrate historical data and domain nuances.

Decisions on open-source versus proprietary models and in-house versus vendor-based deployment also interact with the chosen adaptation approach. An institution that values data sovereignty may prefer an open-source solution augmented by PEFT for cost effectiveness, while one lacking extensive AI resources could opt for a proprietary model supplemented by RAG for continuous factual grounding.

\section{Integrated decision flow for responsible LLM adoption}

As emphasized in previous discussions, practical operational safeguards translate high-level strategic decisions—such as data governance, risk management, ethical standards, and ROI priorities—into everyday institutional routines (Table~\ref{tab:tech_mitigation_streamlined}). Early and proactive implementation of these safeguards helps prevent costly misalignment with critical mandates on privacy, compliance, or governance. For example, measures like data localization assessments and federated learning strategies often need resolution before committing to vendor-based or on-premises deployment approaches. Similarly, bias audits and explainability requirements begin shaping model configurations early in the decision-making process, prior to finalizing any implementation strategy. Once a deployment path is selected, institutions further refine these safeguards—adopting specific encryption standards for in-house implementations, for example, or relying on vendor-provided compliance certifications.

\begin{table*}[ht]
\centering
\footnotesize
\caption{Key operational safeguards for responsible LLM deployment in finance}
\label{tab:tech_mitigation_streamlined}
\resizebox{\textwidth}{!}{%
\begin{tabular}{p{2.6cm} p{2.4cm} p{4.2cm} p{4.1cm}}
\toprule
\textbf{Method} 
& \textbf{Core Purpose} 
& \textbf{Implementation Essentials} 
& \textbf{If Neglected}\\
\midrule

\textbf{Data Classification \& Minimization} 
& Align with privacy laws \newline (\nameref{decision:2})
& Categorize data by sensitivity; filter out unnecessary fields; ensure compliance (e.g.\ GDPR).
& Sensitive data leaks; major fines; reputational damage\\

\hline
\textbf{Federated Learning} 
& Meet data-localization mandates \newline (\nameref{decision:2})
& Partition datasets regionally; transmit only model parameters; avoid exposing private fields.
& Violations of local rules; consent breaches; enforcement actions\\

\hline

\textbf{Anonymization, Encryption, \& Secure Enclaves} 
& Protect confidential and PII (\nameref{decision:3})
& Tokenize personal info; encrypt pipelines; use secure enclaves; keep records for audits.
& Unencrypted data triggers trust breaches, legal penalties, service suspensions\\

\hline

\textbf{Explainability Tools \& Chain-of-Thought} 
& Meet audit and transparency needs \newline (\nameref{decision:3})
& Integrate LIME/SHAP logs; produce “model cards” detailing biases; store rationale for compliance.
& Opaque decisions; noncompliance with explainability mandates; user mistrust\\

\hline

\textbf{Human-in-the-Loop (HITL) Oversight} 
& Expert review for critical outputs \newline (\nameref{decision:3})
& Define thresholds (large loans, suspicious alerts); require specialist sign-off; log overrides.
& Faulty automation in high-stakes decisions; legal/ethical exposure\\

\hline

\textbf{Incident Response \& Governance Committees} 
& Ongoing oversight; address drift or breaches (\nameref{decision:3})
& Formal escalation plans; cross-functional committees; regularly revise policies .
& Delayed breach handling; regulatory fallout; erosion of public trust\\
\hline
\textbf{Bias Auditing \& Debiasing} 
& Maintain fairness; comply with FEAT \newline (\nameref{decision:4A})
& Regular bias checks; scenario testing; adjust data or apply fairness constraints if disparities exist.
& Undetected discrimination; fair-lending violations; reputational harm\\

\hline

\textbf{Sandbox/Pilot Testing \& Adversarial Prompts} 
& Validate compliance pre-deployment
& Run controlled pilots; stress-test with malicious inputs; fix vulnerabilities before scale-up.
& Large-scale failures; negative user backlash; post-launch compliance breaches\\

\bottomrule
\end{tabular}
}
\end{table*}

\begin{figure}[h!] 
  \centering
  \fbox{\includegraphics[width=0.5\textwidth]{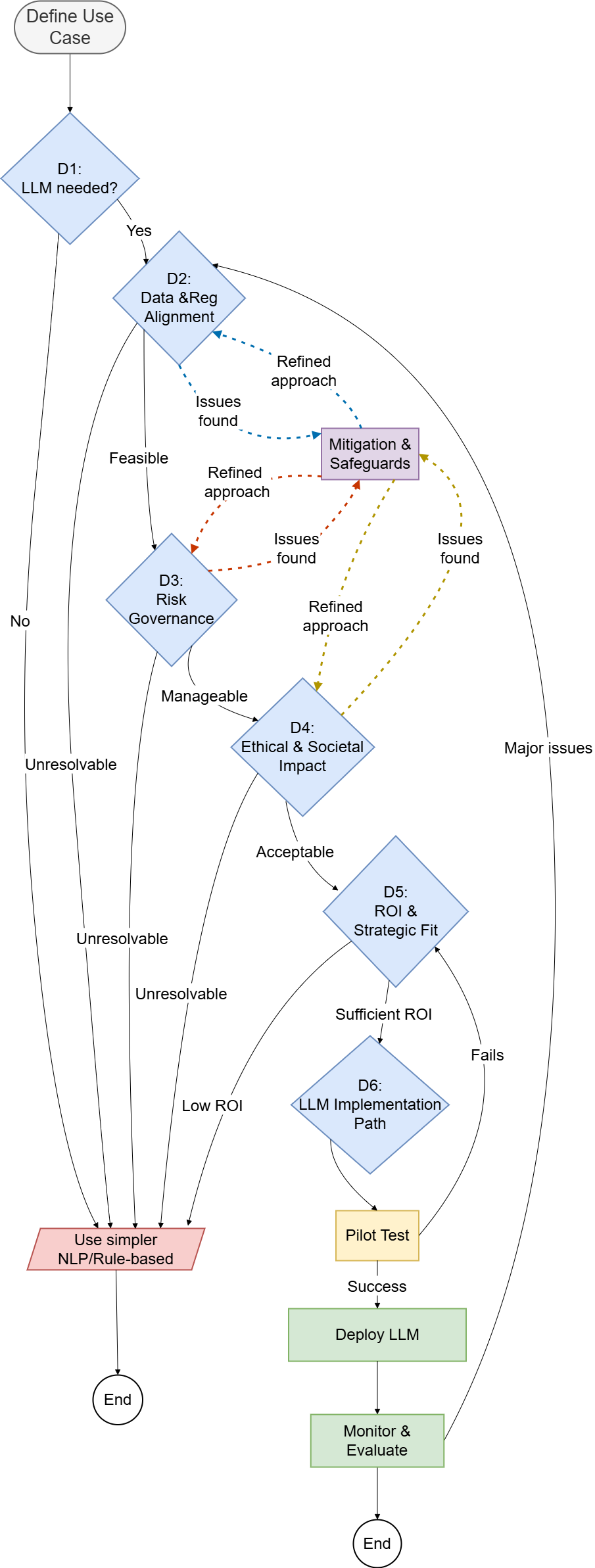}}  
  \caption{Decision flow for responsible LLM innovation in finance.}
  \label{fig:decision1-6_flowchart} 
\end{figure}

Figure~\ref{fig:decision1-6_flowchart} synthesizes all six key decision points into a unified workflow. It begins by determining whether an advanced LLM is genuinely necessary (\nameref{decision:1}), then moves through data feasibility (\nameref{decision:2}), governance readiness (\nameref{decision:3}), ethical considerations (\nameref{decision:4A}), and strategic ROI (\nameref{decision:4}). Upon meeting these criteria, the institution finalizes an implementation plan (\nameref{decision:6}) and conducts a pilot test to confirm initial assumptions. Any shortfalls during the pilot—such as insufficient ROI, privacy conflicts, or security gaps—lead back to earlier decisions for remediation.

Once in production, a continuous “monitor and evaluate” phase follows, relying on the metrics in Table~\ref{tab:strategic_llm_evaluation} (e.g., performance, trust and safety, operational efficiency, user experience, and risk management). \emph{Major} issues—like newly imposed data-localization mandates or systemic biases—prompt a return to \nameref{decision:2} for renewed data checks and upgraded safeguards. \emph{Minor} problems—such as small latency spikes or mild user complaints—are typically addressed with incremental model refinements, updated prompt-engineering strategies, or minor tweaks to oversight procedures.

This iterative and adaptive approach ensures early-stage decisions remain responsive to changing operational contexts. Continuous feedback enables institutions to swiftly address new compliance demands and user needs, maintaining alignment with privacy, security, fairness, and strategic business goals throughout the LLM lifecycle—from pilot testing through mature operations.

\begin{table}[ht]
\centering
\footnotesize
\caption{Key metrics for LLM monitoring and evaluation}
\label{tab:strategic_llm_evaluation}
\begin{tabular}{p{0.15\linewidth} p{0.37\linewidth} p{0.38\linewidth}}
\toprule
\textbf{Dimension} & \textbf{Key Metrics} & \textbf{Strategic Importance} \\
\midrule
\textbf{Performance}
& Accuracy, recall, domain coverage, response coherence
& Validates whether the LLM delivers consistent value in core tasks, reinforcing its competitive viability \\
\midrule
\textbf{Trust \& Safety}
& Factual accuracy, fairness indices, content moderation
& Preserves stakeholder and regulatory confidence, minimizing reputational risks \\
\midrule
\textbf{Operational Efficiency}
& Latency, throughput, resource usage, scaling costs
& Governs cost-effectiveness and seamless user experiences under varied load \\
\midrule
\textbf{User Experience}
& Satisfaction scores, task completion rates, engagement levels
& Encourages adoption and long-term utility through positive end-user outcomes \\
\midrule
\textbf{Risk Management}
& Security monitoring, privacy compliance, anomaly detection
& Safeguards institutional resilience, aligning with governance and oversight frameworks \\
\bottomrule
\end{tabular}
\end{table}

\section{Conclusion}
Integrating advanced LLMs into finance can position institutions at the intersection of sophisticated linguistic capabilities and rigorous data-driven analysis. However, realizing this potential necessitates careful navigation of complex regulatory environments, ethical implications, and consequential financial decisions. Our analysis of the six key decisions—determining the necessity of LLMs, ensuring robust data governance, managing associated risks, instituting ethical oversight, systematically evaluating ROI, and selecting optimal deployment strategies—emphasizes the importance of embedding AI ambitions into practical operational and strategic realities.

Institutions positioned to thrive will adopt a disciplined approach, thoughtfully balancing innovative generative AI capabilities against critical requirements for transparency, compliance, and financial prudence. Early pilot projects are essential for identifying vulnerabilities, such as unintended biases or data privacy exposures. Effective integration will particularly rely on hybrid workflows, where AI complements and enhances professional expertise without fully automating sensitive financial judgments, thus preserving stakeholder trust and maintaining accountability.

Moving forward, adaptability will be essential. Institutions must continuously revisit and refine the proposed framework, responding proactively to changing regulatory conditions, emerging ethical guidelines, and advances in AI architectures. Broader strategic challenges, such as reconciling divergent international regulations and managing biases revealed by AI systems, will increasingly shape the adoption trajectory.

Ultimately, the successful adoption of LLMs in finance represents a balanced combination of technological advancement, rigorous governance, and ethical vigilance. The decision framework proposed in this paper serves as a flexible, ongoing guide that emphasizes transparency, accountability, and iterative improvement. By aligning with these principles, financial institutions can better leverage LLMs not merely as tools for operational efficiency but as instruments for delivering more transparent, accountable, and ethically sound financial services, while reinforcing institutional integrity and public confidence.

\bibliography{reference.bib}

\begin{thebibliography}{71}
\providecommand{\natexlab}[1]{#1}
\providecommand{\url}[1]{\texttt{#1}}
\expandafter\ifx\csname urlstyle\endcsname\relax
  \providecommand{\doi}[1]{doi: #1}\else
  \providecommand{\doi}{doi: \begingroup \urlstyle{rm}\Url}\fi

\bibitem[Araci(2019)]{araci_finbert:_2019}
D.~Araci.
\newblock {FinBERT}: {Financial} {Sentiment} {Analysis} with {Pre}-trained {Language} {Models}.
\newblock arXiv:1908.10063, Aug. 2019.
\newblock URL \url{http://arxiv.org/abs/1908.10063}.

\bibitem[Askell et~al.(2021)Askell, Bai, Chen, Drain, Ganguli, Henighan, et~al.]{askell_general_2021}
A.~Askell, Y.~Bai, A.~Chen, D.~Drain, D.~Ganguli, T.~Henighan, et~al.
\newblock A {General} {Language} {Assistant} as a {Laboratory} for {Alignment}.
\newblock arXiv:2112.00861, Dec. 2021.
\newblock URL \url{http://arxiv.org/abs/2112.00861}.

\bibitem[Bakumenko et~al.(2024)Bakumenko, Hlaváčková-Schindler, Plant, and Hubig]{bakumenko_advancing_2024}
A.~Bakumenko, K.~Hlaváčková-Schindler, C.~Plant, and N.~C. Hubig.
\newblock Advancing {Anomaly} {Detection}: {Non}-{Semantic} {Financial} {Data} {Encoding} with {LLMs}.
\newblock arXiv:2406.03614, June 2024.
\newblock URL \url{http://arxiv.org/abs/2406.03614}.

\bibitem[Barocas and Selbst(2016)]{barocas_big_2016}
S.~Barocas and A.~D. Selbst.
\newblock Big {Data}'s {Disparate} {Impact}.
\newblock \emph{SSRN Electronic Journal}, 2016.
\newblock ISSN 1556-5068.
\newblock \doi{10.2139/ssrn.2477899}.
\newblock URL \url{https://www.ssrn.com/abstract=2477899}.

\bibitem[{Basel Committee on Banking Supervision}(2018)]{bcbs2018}
{Basel Committee on Banking Supervision}.
\newblock {Sound Practices: Implications of Fintech Developments for Banks and Bank Supervisors}.
\newblock Technical report, Bank for International Settlements, 2018.

\bibitem[Bommasani et~al.(2022)Bommasani, Hudson, Adeli, Altman, Arora, Arx, et~al.]{bommasani_opportunities_2022}
R.~Bommasani, D.~A. Hudson, E.~Adeli, R.~Altman, S.~Arora, S.~v. Arx, et~al.
\newblock On the {Opportunities} and {Risks} of {Foundation} {Models}.
\newblock arXiv:2108.07258, July 2022.
\newblock URL \url{http://arxiv.org/abs/2108.07258}.

\bibitem[Brown et~al.(2020)Brown, Mann, Ryder, Subbiah, Kaplan, Dhariwal, et~al.]{brown_language_2020}
T.~Brown, B.~Mann, N.~Ryder, M.~Subbiah, J.~D. Kaplan, P.~Dhariwal, et~al.
\newblock Language {Models} are {Few}-{Shot} {Learners}.
\newblock In \emph{Advances in {Neural} {Information} {Processing} {Systems}}, volume~33, pages 1877--1901. Curran Associates, Inc., 2020.
\newblock URL \url{https://proceedings.neurips.cc/paper/2020/hash/1457c0d6bfcb4967418bfb8ac142f64a-Abstract.html}.

\bibitem[Cao and Feinstein(2024)]{cao_large_2024}
Z.~Cao and Z.~Feinstein.
\newblock Large {Language} {Model} in {Financial} {Regulatory} {Interpretation}.
\newblock arXiv:2405.06808, July 2024.
\newblock URL \url{http://arxiv.org/abs/2405.06808}.

\bibitem[Carlini et~al.(2021)Carlini, Tramèr, Wallace, Jagielski, Herbert-Voss, Lee, et~al.]{carlini_extracting_2021}
N.~Carlini, F.~Tramèr, E.~Wallace, M.~Jagielski, A.~Herbert-Voss, K.~Lee, et~al.
\newblock Extracting {Training} {Data} from {Large} {Language} {Models}.
\newblock pages 2633--2650, 2021.
\newblock ISBN 9781939133243.
\newblock URL \url{https://www.usenix.org/conference/usenixsecurity21/presentation/carlini-extracting}.

\bibitem[Chang et~al.(2024)Chang, Wang, Wang, Wu, Yang, Zhu, et~al.]{chang_survey_2024}
Y.~Chang, X.~Wang, J.~Wang, Y.~Wu, L.~Yang, K.~Zhu, et~al.
\newblock A {Survey} on {Evaluation} of {Large} {Language} {Models}.
\newblock \emph{ACM Trans. Intell. Syst. Technol.}, 15\penalty0 (3):\penalty0 39:1--39:45, Mar. 2024.
\newblock ISSN 2157-6904.
\newblock \doi{10.1145/3641289}.
\newblock URL \url{https://dl.acm.org/doi/10.1145/3641289}.

\bibitem[Chapman et~al.(2022)Chapman, Hillebrand, Stenzel, Deußer, Biesner, Bauckhage, et~al.]{chapman_towards_2022}
C.~L. Chapman, L.~Hillebrand, M.~R. Stenzel, T.~Deußer, D.~Biesner, C.~Bauckhage, et~al.
\newblock Towards {Generating} {Financial} {Reports} from {Tabular} {Data} {Using} {Transformers}.
\newblock In A.~Holzinger, P.~Kieseberg, A.~M. Tjoa, and E.~Weippl, editors, \emph{Machine {Learning} and {Knowledge} {Extraction}}, pages 221--232, Cham, 2022. Springer International Publishing.
\newblock ISBN 9783031144639.
\newblock \doi{10.1007/978-3-031-14463-9_14}.

\bibitem[Che et~al.(2023)Che, Liu, Zhou, Ren, Zhou, Sheng, Dai, and Dou]{che2023federated}
T.~Che, J.~Liu, Y.~Zhou, J.~Ren, J.~Zhou, V.~S. Sheng, H.~Dai, and D.~Dou.
\newblock Federated learning of large language models with parameter-efficient prompt tuning and adaptive optimization.
\newblock \emph{arXiv preprint arXiv:2310.15080}, 2023.

\bibitem[Chen et~al.(2023)Chen, Fu, Yuan, Wen, Fan, Liu, et~al.]{chen_hallucination_2023}
Y.~Chen, Q.~Fu, Y.~Yuan, Z.~Wen, G.~Fan, D.~Liu, et~al.
\newblock Hallucination {Detection}: {Robustly} {Discerning} {Reliable} {Answers} in {Large} {Language} {Models}.
\newblock In \emph{Proceedings of the 32nd {ACM} {International} {Conference} on {Information} and {Knowledge} {Management}}, pages 245--255, Birmingham United Kingdom, Oct. 2023. ACM.
\newblock ISBN 9798400701245.
\newblock \doi{10.1145/3583780.3614905}.
\newblock URL \url{https://dl.acm.org/doi/10.1145/3583780.3614905}.

\bibitem[Chen et~al.(2022)Chen, Chen, Smiley, Shah, Borova, Langdon, et~al.]{chen_finqa:_2022}
Z.~Chen, W.~Chen, C.~Smiley, S.~Shah, I.~Borova, D.~Langdon, et~al.
\newblock {FinQA}: {A} {Dataset} of {Numerical} {Reasoning} over {Financial} {Data}.
\newblock arXiv:2109.00122, May 2022.
\newblock URL \url{http://arxiv.org/abs/2109.00122}.

\bibitem[Crisanto et~al.(2024)Crisanto, Leuterio, Prenio, and Yong]{crisanto_regulating_2024}
J.~C. Crisanto, C.~B. Leuterio, J.~Prenio, and J.~Yong.
\newblock Regulating {AI} in the financial sector: recent developments and main challenges.
\newblock Dec. 2024.
\newblock URL \url{https://www.bis.org/fsi/publ/insights63.htm}.

\bibitem[De~Lange et~al.(2022)De~Lange, Aljundi, Masana, Parisot, Jia, Leonardis, et~al.]{de_lange_continual_2022}
M.~De~Lange, R.~Aljundi, M.~Masana, S.~Parisot, X.~Jia, A.~Leonardis, et~al.
\newblock A {Continual} {Learning} {Survey}: {Defying} {Forgetting} in {Classification} {Tasks}.
\newblock \emph{IEEE Transactions on Pattern Analysis and Machine Intelligence}, 44\penalty0 (7):\penalty0 3366--3385, July 2022.
\newblock ISSN 1939-3539.
\newblock \doi{10.1109/TPAMI.2021.3057446}.
\newblock URL \url{https://ieeexplore.ieee.org/abstract/document/9349197/}.

\bibitem[Dettmers et~al.(2023)Dettmers, Pagnoni, Holtzman, and Zettlemoyer]{dettmers_qlora:_2023}
T.~Dettmers, A.~Pagnoni, A.~Holtzman, and L.~Zettlemoyer.
\newblock {QLoRA}: {Efficient} {Finetuning} of {Quantized} {LLMs}.
\newblock \emph{Advances in Neural Information Processing Systems}, 36:\penalty0 10088--10115, Dec. 2023.
\newblock URL \url{https://proceedings.neurips.cc/paper_files/paper/2023/hash/1feb87871436031bdc0f2beaa62a049b-Abstract-Conference.html}.

\bibitem[Devlin et~al.(2019)Devlin, Chang, Lee, and Toutanova]{devlin_bert:_2019}
J.~Devlin, M.-W. Chang, K.~Lee, and K.~Toutanova.
\newblock {BERT}: {Pre}-training of {Deep} {Bidirectional} {Transformers} for {Language} {Understanding}.
\newblock In J.~Burstein, C.~Doran, and T.~Solorio, editors, \emph{Proceedings of the 2019 {Conference} of the {North} {American} {Chapter} of the {Association} for {Computational} {Linguistics}: {Human} {Language} {Technologies}, {Volume} 1 ({Long} and {Short} {Papers})}, pages 4171--4186, Minneapolis, Minnesota, June 2019. Association for Computational Linguistics.
\newblock \doi{10.18653/v1/N19-1423}.
\newblock URL \url{https://aclanthology.org/N19-1423/}.

\bibitem[{European Banking Authority}(2021)]{eba2021}
{European Banking Authority}.
\newblock {EBA Report on Big Data and Advanced Analytics}.
\newblock Technical report, European Banking Authority, 2021.
\newblock URL \url{https://www.eba.europa.eu}.

\bibitem[Ferrara(2023)]{ferrara_should_2023}
E.~Ferrara.
\newblock Should {ChatGPT} be {Biased}? {Challenges} and {Risks} of {Bias} in {Large} {Language} {Models}.
\newblock arXiv:2304.03738, Nov. 2023.
\newblock URL \url{http://arxiv.org/abs/2304.03738}.

\bibitem[Finn et~al.(2017)Finn, Abbeel, and Levine]{finn_model-agnostic_2017}
C.~Finn, P.~Abbeel, and S.~Levine.
\newblock Model-{Agnostic} {Meta}-{Learning} for {Fast} {Adaptation} of {Deep} {Networks}.
\newblock In \emph{Proceedings of the 34th {International} {Conference} on {Machine} {Learning}}, pages 1126--1135. PMLR, July 2017.
\newblock URL \url{https://proceedings.mlr.press/v70/finn17a.html}.

\bibitem[{GDPR}(2016)]{GDPR2016}
{GDPR}.
\newblock General data protection regulation.
\newblock https://eur-lex.europa.eu/eli/reg/2016/679/oj/eng, 2016.

\bibitem[Gou et~al.(2021)Gou, Yu, Maybank, and Tao]{gou_knowledge_2021}
J.~Gou, B.~Yu, S.~J. Maybank, and D.~Tao.
\newblock Knowledge {Distillation}: {A} {Survey}.
\newblock \emph{International Journal of Computer Vision}, 129\penalty0 (6):\penalty0 1789--1819, June 2021.
\newblock ISSN 1573-1405.
\newblock \doi{10.1007/s11263-021-01453-z}.
\newblock URL \url{https://doi.org/10.1007/s11263-021-01453-z}.

\bibitem[Guidotti et~al.(2018)Guidotti, Monreale, Ruggieri, Turini, Giannotti, and Pedreschi]{guidotti_survey_2018}
R.~Guidotti, A.~Monreale, S.~Ruggieri, F.~Turini, F.~Giannotti, and D.~Pedreschi.
\newblock A {Survey} of {Methods} for {Explaining} {Black} {Box} {Models}.
\newblock \emph{ACM Comput. Surv.}, 51\penalty0 (5):\penalty0 93:1--93:42, Aug. 2018.
\newblock ISSN 0360-0300.
\newblock \doi{10.1145/3236009}.
\newblock URL \url{https://dl.acm.org/doi/10.1145/3236009}.

\bibitem[Gupta(2024)]{gupta_ai_2024}
A.~Gupta.
\newblock The {AI} {Capability} {Continuum}: {A} {Three} {Step} {Framework} for {AI} {System} {Development}, Mar. 2024.
\newblock URL \url{https://papers.ssrn.com/abstract=4754694}.

\bibitem[Han et~al.(2024)Han, Gao, Liu, Zhang, and Zhang]{han_parameter-efficient_2024}
Z.~Han, C.~Gao, J.~Liu, J.~Zhang, and S.~Q. Zhang.
\newblock Parameter-{Efficient} {Fine}-{Tuning} for {Large} {Models}: {A} {Comprehensive} {Survey}.
\newblock arXiv:2403.14608, Sept. 2024.
\newblock URL \url{http://arxiv.org/abs/2403.14608}.

\bibitem[Houlsby et~al.(2019)Houlsby, Giurgiu, Jastrzebski, Morrone, Laroussilhe, Gesmundo, et~al.]{houlsby_parameter-efficient_2019}
N.~Houlsby, A.~Giurgiu, S.~Jastrzebski, B.~Morrone, Q.~D. Laroussilhe, A.~Gesmundo, et~al.
\newblock Parameter-{Efficient} {Transfer} {Learning} for {NLP}.
\newblock In \emph{Proceedings of the 36th {International} {Conference} on {Machine} {Learning}}, pages 2790--2799. PMLR, May 2019.
\newblock URL \url{https://proceedings.mlr.press/v97/houlsby19a.html}.

\bibitem[Hu et~al.(2021)Hu, Shen, Wallis, Allen-Zhu, Li, Wang, et~al.]{hu_lora:_2021}
E.~J. Hu, Y.~Shen, P.~Wallis, Z.~Allen-Zhu, Y.~Li, S.~Wang, et~al.
\newblock {LoRA}: {Low}-{Rank} {Adaptation} of {Large} {Language} {Models}.
\newblock arXiv:2106.09685, Oct. 2021.
\newblock URL \url{http://arxiv.org/abs/2106.09685}.

\bibitem[Hu et~al.(2024)Hu, Qin, Yuan, Peng, Lopez-Lira, Wang, et~al.]{hu_no_2024}
G.~Hu, K.~Qin, C.~Yuan, M.~Peng, A.~Lopez-Lira, B.~Wang, et~al.
\newblock No {Language} is an {Island}: {Unifying} {Chinese} and {English} in {Financial} {Large} {Language} {Models}, {Instruction} {Data}, and {Benchmarks}.
\newblock arXiv:2403.06249, Aug. 2024.
\newblock URL \url{http://arxiv.org/abs/2403.06249}.

\bibitem[Huang et~al.(2023)Huang, Wang, and Yang]{huang_<span_2023}
A.~H. Huang, H.~Wang, and Y.~Yang.
\newblock {FinBERT}: {A} {Large} {Language} {Model} for {Extracting} {Information} from {Financial} {Text}*.
\newblock \emph{Contemporary Accounting Research}, 40\penalty0 (2):\penalty0 806--841, May 2023.
\newblock ISSN 0823-9150, 1911-3846.
\newblock \doi{10.1111/1911-3846.12832}.
\newblock URL \url{https://onlinelibrary.wiley.com/doi/10.1111/1911-3846.12832}.

\bibitem[Irugalbandara et~al.(2024)Irugalbandara, Mahendra, Daynauth, Arachchige, Dantanarayana, Flautner, et~al.]{irugalbandara_scaling_2024}
C.~Irugalbandara, A.~Mahendra, R.~Daynauth, T.~K. Arachchige, J.~Dantanarayana, K.~Flautner, et~al.
\newblock Scaling {Down} to {Scale} {Up}: {A} {Cost}-{Benefit} {Analysis} of {Replacing} {OpenAI}'s {LLM} with {Open} {Source} {SLMs} in {Production}.
\newblock arXiv:2312.14972, Apr. 2024.
\newblock URL \url{http://arxiv.org/abs/2312.14972}.

\bibitem[Ji et~al.(2023)Ji, Lee, Frieske, Yu, Su, Xu, et~al.]{ji_survey_2023}
Z.~Ji, N.~Lee, R.~Frieske, T.~Yu, D.~Su, Y.~Xu, et~al.
\newblock Survey of {Hallucination} in {Natural} {Language} {Generation}.
\newblock \emph{ACM Comput. Surv.}, 55\penalty0 (12):\penalty0 248:1--248:38, Mar. 2023.
\newblock ISSN 0360-0300.
\newblock \doi{10.1145/3571730}.
\newblock URL \url{https://dl.acm.org/doi/10.1145/3571730}.

\bibitem[Kirk et~al.(2024)Kirk, Mediratta, Nalmpantis, Luketina, Hambro, Grefenstette, et~al.]{kirk_understanding_2024}
R.~Kirk, I.~Mediratta, C.~Nalmpantis, J.~Luketina, E.~Hambro, E.~Grefenstette, et~al.
\newblock Understanding the {Effects} of {RLHF} on {LLM} {Generalisation} and {Diversity}.
\newblock arXiv:2310.06452, Feb. 2024.
\newblock URL \url{http://arxiv.org/abs/2310.06452}.

\bibitem[Lakkaraju et~al.(2023)Lakkaraju, Jones, Vuruma, Pallagani, Muppasani, and Srivastava]{lakkaraju_llms_2023}
K.~Lakkaraju, S.~E. Jones, S.~K.~R. Vuruma, V.~Pallagani, B.~C. Muppasani, and B.~Srivastava.
\newblock {LLMs} for {Financial} {Advisement}: {A} {Fairness} and {Efficacy} {Study} in {Personal} {Decision} {Making}.
\newblock In \emph{Proceedings of the {Fourth} {ACM} {International} {Conference} on {AI} in {Finance}}, {ICAIF} '23, pages 100--107, New York, NY, USA, Nov. 2023. Association for Computing Machinery.
\newblock ISBN 9798400702402.
\newblock \doi{10.1145/3604237.3626867}.
\newblock URL \url{https://dl.acm.org/doi/10.1145/3604237.3626867}.

\bibitem[Langedijk et~al.(2022)Langedijk, Dankers, Lippe, Bos, Cardenas~Guevara, Yannakoudakis, et~al.]{langedijk_meta-learning_2022}
A.~Langedijk, V.~Dankers, P.~Lippe, S.~Bos, B.~Cardenas~Guevara, H.~Yannakoudakis, et~al.
\newblock Meta-{Learning} for {Fast} {Cross}-{Lingual} {Adaptation} in {Dependency} {Parsing}.
\newblock In \emph{Proceedings of the 60th {Annual} {Meeting} of the {Association} for {Computational} {Linguistics} ({Volume} 1: {Long} {Papers})}, pages 8503--8520, Dublin, Ireland, 2022. Association for Computational Linguistics.
\newblock \doi{10.18653/v1/2022.acl-long.582}.
\newblock URL \url{https://aclanthology.org/2022.acl-long.582}.

\bibitem[Lee et~al.(2025)Lee, Stevens, and Han]{lee_large_2025}
J.~Lee, N.~Stevens, and S.~C. Han.
\newblock Large {Language} {Models} in {Finance} ({FinLLMs}).
\newblock \emph{Neural Computing and Applications}, Jan. 2025.
\newblock ISSN 1433-3058.
\newblock \doi{10.1007/s00521-024-10495-6}.
\newblock URL \url{https://doi.org/10.1007/s00521-024-10495-6}.

\bibitem[Lewis et~al.(2020)Lewis, Perez, Piktus, Petroni, Karpukhin, Goyal, Küttler, et~al.]{lewis_retrieval-augmented_2020}
P.~Lewis, E.~Perez, A.~Piktus, F.~Petroni, V.~Karpukhin, N.~Goyal, H.~Küttler, et~al.
\newblock Retrieval-{Augmented} {Generation} for {Knowledge}-{Intensive} {NLP} {Tasks}.
\newblock In \emph{Advances in {Neural} {Information} {Processing} {Systems}}, volume~33, pages 9459--9474. Curran Associates, Inc., 2020.
\newblock URL \url{https://proceedings.neurips.cc/paper/2020/hash/6b493230205f780e1bc26945df7481e5-Abstract.html}.

\bibitem[Li and Liang(2021)]{li_prefix-tuning:_2021}
X.~L. Li and P.~Liang.
\newblock Prefix-{Tuning}: {Optimizing} {Continuous} {Prompts} for {Generation}.
\newblock arXiv:2101.00190, Jan. 2021.
\newblock URL \url{http://arxiv.org/abs/2101.00190}.

\bibitem[Li et~al.(2023)Li, Wang, Ding, and Chen]{li_large_2023}
Y.~Li, S.~Wang, H.~Ding, and H.~Chen.
\newblock Large {Language} {Models} in {Finance}: {A} {Survey}.
\newblock In \emph{Proceedings of the {Fourth} {ACM} {International} {Conference} on {AI} in {Finance}}, {ICAIF} '23, pages 374--382, New York, NY, USA, Nov. 2023. Association for Computing Machinery.
\newblock ISBN 9798400702402.
\newblock \doi{10.1145/3604237.3626869}.
\newblock URL \url{https://dl.acm.org/doi/10.1145/3604237.3626869}.

\bibitem[Li et~al.(2024)Li, Wang, Ding, and Chen]{li_large_2024}
Y.~Li, S.~Wang, H.~Ding, and H.~Chen.
\newblock Large {Language} {Models} in {Finance}: {A} {Survey}.
\newblock arXiv:2311.10723, July 2024.
\newblock URL \url{http://arxiv.org/abs/2311.10723}.

\bibitem[Liang et~al.(2021)Liang, Wu, Morency, and Salakhutdinov]{liang_towards_2021}
P.~P. Liang, C.~Wu, L.-P. Morency, and R.~Salakhutdinov.
\newblock Towards {Understanding} and {Mitigating} {Social} {Biases} in {Language} {Models}.
\newblock In \emph{Proceedings of the 38th {International} {Conference} on {Machine} {Learning}}, pages 6565--6576. PMLR, July 2021.
\newblock URL \url{https://proceedings.mlr.press/v139/liang21a.html}.

\bibitem[Liu et~al.(2024)Liu, Yao, Ton, Zhang, Guo, Cheng, et~al.]{liu_trustworthy_2024}
Y.~Liu, Y.~Yao, J.-F. Ton, X.~Zhang, R.~Guo, H.~Cheng, et~al.
\newblock Trustworthy {LLMs}: a {Survey} and {Guideline} for {Evaluating} {Large} {Language} {Models}' {Alignment}.
\newblock arXiv:2308.05374, Mar. 2024.
\newblock URL \url{http://arxiv.org/abs/2308.05374}.

\bibitem[Lo and Ross(2024)]{lo_can_2024}
A.~W. Lo and J.~Ross.
\newblock Can {ChatGPT} {Plan} {Your} {Retirement}?: {Generative} {AI} and {Financial} {Advice}.
\newblock \emph{Harvard Data Science Review}, \penalty0 (Special Issue 5), Aug. 2024.
\newblock ISSN 2644-2353, 2688-8513.
\newblock \doi{10.1162/99608f92.ec74a002}.
\newblock URL \url{https://hdsr.mitpress.mit.edu/pub/jnml28pl/release/3}.

\bibitem[Long et~al.(2020{\natexlab{a}})Long, Tan, Jiang, and Zhang]{long2020federated}
G.~Long, Y.~Tan, J.~Jiang, and C.~Zhang.
\newblock Federated learning for open banking.
\newblock In \emph{Federated learning: privacy and incentive}, pages 240--254. Springer, 2020{\natexlab{a}}.

\bibitem[Long et~al.(2020{\natexlab{b}})Long, Tan, Jiang, and Zhang]{long_federated_2020}
G.~Long, Y.~Tan, J.~Jiang, and C.~Zhang.
\newblock Federated {Learning} for {Open} {Banking}.
\newblock In Q.~Yang, L.~Fan, and H.~Yu, editors, \emph{Federated {Learning}: {Privacy} and {Incentive}}, pages 240--254. Springer International Publishing, Cham, 2020{\natexlab{b}}.
\newblock ISBN 9783030630768.
\newblock \doi{10.1007/978-3-030-63076-8_17}.
\newblock URL \url{https://doi.org/10.1007/978-3-030-63076-8_17}.

\bibitem[McCarthy(2022)]{mccarthy_regulation_2022}
J.~McCarthy.
\newblock The regulation of {RegTech} and {SupTech} in finance: ensuring consistency in principle and in practice.
\newblock \emph{Journal of Financial Regulation and Compliance}, 31\penalty0 (2):\penalty0 186--199, Jan. 2022.
\newblock ISSN 1358-1988.
\newblock \doi{10.1108/JFRC-01-2022-0004}.
\newblock URL \url{https://doi.org/10.1108/JFRC-01-2022-0004}.

\bibitem[Mehrabi et~al.(2022)Mehrabi, Morstatter, Saxena, Lerman, and Galstyan]{mehrabi_survey_2022}
N.~Mehrabi, F.~Morstatter, N.~Saxena, K.~Lerman, and A.~Galstyan.
\newblock A {Survey} on {Bias} and {Fairness} in {Machine} {Learning}.
\newblock \emph{ACM Computing Surveys}, 54\penalty0 (6):\penalty0 1--35, July 2022.
\newblock ISSN 0360-0300, 1557-7341.
\newblock \doi{10.1145/3457607}.
\newblock URL \url{https://dl.acm.org/doi/10.1145/3457607}.

\bibitem[Mökander et~al.(2024)Mökander, Schuett, Kirk, and Floridi]{mokander_auditing_2024-1}
J.~Mökander, J.~Schuett, H.~R. Kirk, and L.~Floridi.
\newblock Auditing large language models: a three-layered approach.
\newblock \emph{AI and Ethics}, 4\penalty0 (4):\penalty0 1085--1115, Nov. 2024.
\newblock ISSN 2730-5961.
\newblock \doi{10.1007/s43681-023-00289-2}.
\newblock URL \url{https://doi.org/10.1007/s43681-023-00289-2}.

\bibitem[{OpenAI} et~al.(2024){OpenAI}, Achiam, Adler, Agarwal, Ahmad, Akkaya, Aleman, et~al.]{openai_gpt-4_2024}
{OpenAI}, J.~Achiam, S.~Adler, S.~Agarwal, L.~Ahmad, I.~Akkaya, F.~L. Aleman, et~al.
\newblock {GPT}-4 {Technical} {Report}.
\newblock arXiv:2303.08774, Mar. 2024.
\newblock URL \url{http://arxiv.org/abs/2303.08774}.

\bibitem[Peng et~al.(2021)Peng, Chersoni, Hsu, and Huang]{peng_is_2021}
B.~Peng, E.~Chersoni, Y.~Y. Hsu, and C.~R. Huang.
\newblock Is domain adaptation worth your investment? {Comparing} {BERT} and {FinBERT} on financial tasks.
\newblock pages 37--44. Association for Computational Linguistics (ACL), 2021.
\newblock ISBN 9781954085848.
\newblock \doi{10.18653/v1/2021.econlp-1.5}.
\newblock URL \url{http://ira.lib.polyu.edu.hk/handle/10397/91924}.

\bibitem[Pfeiffer et~al.(2020)Pfeiffer, Rücklé, Poth, Kamath, Vulić, Ruder, et~al.]{pfeiffer_adapterhub:_2020}
J.~Pfeiffer, A.~Rücklé, C.~Poth, A.~Kamath, I.~Vulić, S.~Ruder, et~al.
\newblock {AdapterHub}: {A} {Framework} for {Adapting} {Transformers}.
\newblock arXiv:2007.07779, Oct. 2020.
\newblock URL \url{http://arxiv.org/abs/2007.07779}.

\bibitem[Pfeiffer et~al.(2021)Pfeiffer, Kamath, Rücklé, Cho, and Gurevych]{pfeiffer_adapterfusion:_2021}
J.~Pfeiffer, A.~Kamath, A.~Rücklé, K.~Cho, and I.~Gurevych.
\newblock {AdapterFusion}: {Non}-{Destructive} {Task} {Composition} for {Transfer} {Learning}.
\newblock arXiv:2005.00247, Jan. 2021.
\newblock URL \url{http://arxiv.org/abs/2005.00247}.

\bibitem[Rho et~al.(2024)Rho, Kim, Park, Kim, Chae, Cheon, et~al.]{rho2024encryption}
D.~Rho, T.~Kim, M.~Park, J.~W. Kim, H.~Chae, J.~H. Cheon, et~al.
\newblock Encryption-friendly llm architecture.
\newblock \emph{arXiv preprint arXiv:2410.02486}, 2024.

\bibitem[Ross et~al.(2024)Ross, Kim, and Lo]{ross_llm_2024}
J.~Ross, Y.~Kim, and A.~W. Lo.
\newblock {LLM} economicus? {Mapping} the {Behavioral} {Biases} of {LLMs} via {Utility} {Theory}.
\newblock arXiv:2408.02784, Aug. 2024.
\newblock URL \url{http://arxiv.org/abs/2408.02784}.

\bibitem[Rudin(2019)]{rudin_stop_2019}
C.~Rudin.
\newblock Stop explaining black box machine learning models for high stakes decisions and use interpretable models instead.
\newblock \emph{Nature Machine Intelligence}, 1\penalty0 (5):\penalty0 206--215, May 2019.
\newblock ISSN 2522-5839.
\newblock \doi{10.1038/s42256-019-0048-x}.
\newblock URL \url{https://www.nature.com/articles/s42256-019-0048-x}.

\bibitem[Sanh et~al.(2020)Sanh, Debut, Chaumond, and Wolf]{sanh_distilbert_2020}
V.~Sanh, L.~Debut, J.~Chaumond, and T.~Wolf.
\newblock {DistilBERT}, a distilled version of {BERT}: smaller, faster, cheaper and lighter.
\newblock arXiv:1910.01108, Mar. 2020.
\newblock URL \url{http://arxiv.org/abs/1910.01108}.

\bibitem[Shekhar et~al.(2024)Shekhar, Dubey, Mukherjee, Saxena, Tyagi, and Kotla]{shekhar_towards_2024}
S.~Shekhar, T.~Dubey, K.~Mukherjee, A.~Saxena, A.~Tyagi, and N.~Kotla.
\newblock Towards {Optimizing} the {Costs} of {LLM} {Usage}.
\newblock arXiv:2402.01742, Jan. 2024.
\newblock URL \url{http://arxiv.org/abs/2402.01742}.

\bibitem[Strubell et~al.(2020)Strubell, Ganesh, and McCallum]{strubell_energy_2020}
E.~Strubell, A.~Ganesh, and A.~McCallum.
\newblock Energy and {Policy} {Considerations} for {Modern} {Deep} {Learning} {Research}.
\newblock \emph{Proceedings of the AAAI Conference on Artificial Intelligence}, 34\penalty0 (09):\penalty0 13693--13696, Apr. 2020.
\newblock ISSN 2374-3468.
\newblock \doi{10.1609/aaai.v34i09.7123}.
\newblock URL \url{https://ojs.aaai.org/index.php/AAAI/article/view/7123}.

\bibitem[Wallace et~al.(2021)Wallace, Feng, Kandpal, Gardner, and Singh]{wallace_universal_2021}
E.~Wallace, S.~Feng, N.~Kandpal, M.~Gardner, and S.~Singh.
\newblock Universal {Adversarial} {Triggers} for {Attacking} and {Analyzing} {NLP}.
\newblock arXiv:1908.07125, Jan. 2021.
\newblock URL \url{http://arxiv.org/abs/1908.07125}.

\bibitem[Wei et~al.(2023)Wei, Wang, Schuurmans, Bosma, Ichter, Xia, et~al.]{wei_chain--thought_2023}
J.~Wei, X.~Wang, D.~Schuurmans, M.~Bosma, B.~Ichter, F.~Xia, et~al.
\newblock Chain-of-{Thought} {Prompting} {Elicits} {Reasoning} in {Large} {Language} {Models}.
\newblock arXiv:2201.11903, Jan. 2023.
\newblock URL \url{http://arxiv.org/abs/2201.11903}.

\bibitem[Wu et~al.(2023)Wu, Irsoy, Lu, Dabravolski, Dredze, Gehrmann, et~al.]{wu_bloomberggpt:_2023}
S.~Wu, O.~Irsoy, S.~Lu, V.~Dabravolski, M.~Dredze, S.~Gehrmann, et~al.
\newblock {BloombergGPT}: {A} {Large} {Language} {Model} for {Finance}.
\newblock arXiv:2303.17564, Dec. 2023.
\newblock URL \url{http://arxiv.org/abs/2303.17564}.

\bibitem[Xexéo et~al.(2024)Xexéo, Braida, Parreiras, and Xavier]{xexeo_economic_2024}
G.~Xexéo, F.~Braida, M.~Parreiras, and P.~Xavier.
\newblock The {Economic} {Implications} of {Large} {Language} {Model} {Selection} on {Earnings} and {Return} on {Investment}: {A} {Decision} {Theoretic} {Model}.
\newblock arXiv:2405.17637, May 2024.
\newblock URL \url{http://arxiv.org/abs/2405.17637}.

\bibitem[Xie et~al.(2023)Xie, Han, Zhang, Lai, Peng, Lopez-Lira, and Huang]{xie_pixiu:_2023}
Q.~Xie, W.~Han, X.~Zhang, Y.~Lai, M.~Peng, A.~Lopez-Lira, and J.~Huang.
\newblock {PIXIU}: {A} {Large} {Language} {Model}, {Instruction} {Data} and {Evaluation} {Benchmark} for {Finance}.
\newblock arXiv:2306.05443, June 2023.
\newblock URL \url{http://arxiv.org/abs/2306.05443}.

\bibitem[Xie et~al.(2024)Xie, Li, Xiao, Jiang, Xiang, Zhang, Chen, et~al.]{xie_open-finllms:_2024}
Q.~Xie, D.~Li, M.~Xiao, Z.~Jiang, R.~Xiang, X.~Zhang, Z.~Chen, et~al.
\newblock Open-{FinLLMs}: {Open} {Multimodal} {Large} {Language} {Models} for {Financial} {Applications}.
\newblock arXiv:2408.11878, Aug. 2024.
\newblock URL \url{http://arxiv.org/abs/2408.11878}.

\bibitem[Xie et~al.(2025)Xie, Han, Chen, Xiang, Zhang, He, et~al.]{xie_finben:_2025}
Q.~Xie, W.~Han, Z.~Chen, R.~Xiang, X.~Zhang, Y.~He, et~al.
\newblock {FinBen}: {A} {Holistic} {Financial} {Benchmark} for {Large} {Language} {Models}.
\newblock \emph{Advances in Neural Information Processing Systems}, 37:\penalty0 95716--95743, Jan. 2025.
\newblock URL \url{https://proceedings.neurips.cc/paper_files/paper/2024/hash/adb1d9fa8be4576d28703b396b82ba1b-Abstract-Datasets_and_Benchmarks_Track.html}.

\bibitem[Yang et~al.(2023{\natexlab{a}})Yang, Liu, and Wang]{yang_fingpt:_2023}
H.~Yang, X.-Y. Liu, and C.~D. Wang.
\newblock {FinGPT}: {Open}-{Source} {Financial} {Large} {Language} {Models}.
\newblock arXiv:2306.06031, June 2023{\natexlab{a}}.
\newblock URL \url{http://arxiv.org/abs/2306.06031}.

\bibitem[Yang et~al.(2023{\natexlab{b}})Yang, Wei, Wei, Chen, Huang, Li, et~al.]{yang_survey_2023}
W.~Yang, Y.~Wei, H.~Wei, Y.~Chen, G.~Huang, X.~Li, et~al.
\newblock Survey on {Explainable} {AI}: {From} {Approaches}, {Limitations} and {Applications} {Aspects}.
\newblock \emph{Human-Centric Intelligent Systems}, 3\penalty0 (3):\penalty0 161--188, Sept. 2023{\natexlab{b}}.
\newblock ISSN 2667-1336.
\newblock \doi{10.1007/s44230-023-00038-y}.
\newblock URL \url{https://doi.org/10.1007/s44230-023-00038-y}.

\bibitem[Yang et~al.(2020)Yang, UY, and Huang]{yang_finbert:_2020}
Y.~Yang, M.~C.~S. UY, and A.~Huang.
\newblock {FinBERT}: {A} {Pretrained} {Language} {Model} for {Financial} {Communications}.
\newblock arXiv:2006.08097, July 2020.
\newblock URL \url{http://arxiv.org/abs/2006.08097}.

\bibitem[Yao et~al.(2024)Yao, Zhang, Wu, Huang, Xia, Yu, et~al.]{yao_federated_2024}
Y.~Yao, J.~Zhang, J.~Wu, C.~Huang, Y.~Xia, T.~Yu, et~al.
\newblock Federated {Large} {Language} {Models}: {Current} {Progress} and {Future} {Directions}.
\newblock arXiv:2409.15723, Sept. 2024.
\newblock URL \url{http://arxiv.org/abs/2409.15723}.

\bibitem[Zaken et~al.(2022)Zaken, Ravfogel, and Goldberg]{zaken_bitfit:_2022}
E.~B. Zaken, S.~Ravfogel, and Y.~Goldberg.
\newblock {BitFit}: {Simple} {Parameter}-efficient {Fine}-tuning for {Transformer}-based {Masked} {Language}-models.
\newblock arXiv:2106.10199, Sept. 2022.
\newblock URL \url{http://arxiv.org/abs/2106.10199}.

\bibitem[Zheng et~al.(2023)Zheng, Chiang, Sheng, Zhuang, Wu, Zhuang, et~al.]{zheng_judging_2023}
L.~Zheng, W.-L. Chiang, Y.~Sheng, S.~Zhuang, Z.~Wu, Y.~Zhuang, et~al.
\newblock Judging {LLM}-as-a-{Judge} with {MT}-{Bench} and {Chatbot} {Arena}.
\newblock \emph{Advances in Neural Information Processing Systems}, 36:\penalty0 46595--46623, Dec. 2023.
\newblock URL \url{https://proceedings.neurips.cc/paper_files/paper/2023/hash/91f18a1287b398d378ef22505bf41832-Abstract-Datasets_and_Benchmarks.html}.

\end{thebibliography}
\bibliographystyle{abbrvnat}

\end{document}